# Uniaxial stress flips the natural quantization axis of a quantum dot for integrated quantum photonics

Xueyong Yuan[1], Fritz Weyhausen-Brinkmann[2], Javier Martín-Sánchez[1,3], Giovanni Piredda[4], Vlastimil Křápek[5], Yongheng Huo[1,6,7], Huiying Huang[1], Christian Schimpf[1], Oliver G. Schmidt[6], Johannes Edlinger[4], Gabriel Bester[2], Rinaldo Trotta[1,8], Armando Rastelli*[1]

[1] Institute of Semiconductor and Solid State Physics, Johannes Kepler University Linz, Altenbergerstraße 69, 4040, Linz, Austria

[2] Institut für Physikalische Chemie, Universität Hamburg, Grindelallee 117, 20146 Hamburg, Germany

[3] Departamento de Física, Universidad de Oviedo, 33007 Oviedo, Spain

[4] Forschungszentrum Mikrotechnik, FH Vorarlberg, Hochschulstraße 1, 6850 Dornbirn, Austria

[5] Central European Institute of Technology, Brno University of Technology, Purkyňova 123, 61200 Brno, Czech Republic

[6] Institute for Integrative Nanosciences, IFW Dresden, Helmholtzstraße 20, 01069 Dresden, Germany

[7] Hefei National Laboratory for Physical Sciences at Microscale, University of Science and Technology of China, Shanghai Branch, Xiupu Road 99, 201315 Shanghai, China

[8] Department of Physics, Sapienza University of Rome, Piazzale Aldo Moro 5, 00185 Rome, Italy

*corresponding author: armando.rastelli@jku.at

## Abstract

The optical selection rules in epitaxial quantum dots are strongly influenced by the orientation of their natural quantization axis, which is usually parallel to the growth direction. This configuration is well suited for vertically-emitting devices, but not for planar photonic circuits because of the poorly controlled orientation of the transition dipoles in the growth plane. Here we show that the quantization axis of gallium arsenide dots can be flipped into the growth plane via moderate in-plane uniaxial stress. By using piezoelectric strain-actuators featuring strain-amplification we study the evolution of the selection rules and excitonic fine-structure in a regime, in which quantum confinement can be regarded as a perturbation compared to strain in determining the symmetry-properties of the system. The experimental and computational results suggest that uniaxial stress –may be the right tool to obtain quantum-light sources with ideally oriented transition dipoles and enhanced oscillator strengths for integrated quantum photonics.

# Introduction

Semiconductor quantum dots (QDs) obtained by epitaxial growth are regarded as one of the most promising solid-state sources of triggered single and entangled photons for applications in emerging quantum technologies[1–4]. For this purpose, the radiative recombination of excitons consisting of electrons and holes occupying the lowest energy states confined in the conduction bands (CBs) and valence bands (VBs) of the semiconductor are employed. The selection rules for such transitions in QDs made of common direct-bandgap semiconductors are governed by the nature of the confined holes' ground-state (HGS). In turn, the HGS depends on the confinement potential defined by the structural properties of the QD and surrounding barrier. Epitaxial QDs usually possess flat morphologies, and heights comparable to the Bohr-radius of the confined excitons. Carriers are therefore strongly confined along the growth ($z$) direction, which also defines the natural quantization axis for the total angular momentum operator for the VB states[3–5]. The vertical confinement splits the heavy-hole (HH) and light-hole (LH) bands, so that the HGS has dominant $HH_z$ character, with total angular momentum projection $J_z=\pm 3/2$ (in units of $\hbar$). Dipole-allowed transitions involving such states are characterized by transition-dipoles perpendicular to $z$, making them well suited for efficient vertically-emitting single-photon devices[6–11]. For planar integrated quantum photonics applications[4,12–18], it would be instead desirable to have QDs with transition dipoles perpendicular to the propagation direction, and hence a quantization axis with well-defined orientation in the $x$-$y$ plane. In fact the azimuthal orientation of the transition dipoles for $HH_z$–excitons is usually affected by random fluctuations[19,20], preventing their optimal coupling to guided modes.

In spite of their importance, very little effort has been devoted to developing quantum sources optimized for planar photonic circuits. An exception is represented by Ref.[21], where nanowires-QDs were removed from the substrate and turned by 90° for efficient coupling to dielectric waveguides. Here we show how the natural quantization axis of a QD can be turned to lie in the growth plane without actually rotating the semiconductor matrix, thus preserving the compatibility of the QD heterostructure with planar photonic processing. To this aim we use two key ingredients: (1) high-quality, initially unstrained GaAs QDs[22–24] with a relatively large height, see Fig. 1**a**; (2) in-plane uniaxial stress, provided by micromachined piezoelectric actuators featuring geometric strain amplification. Different from most previous experiments, in which stress was added after growth as a perturbation to fine tune the emission properties of QDs[25,26], confinement can be seen here as a perturbation compared to the strain-induced effects in determining the orientation of the quantization axis and hence the optical selection rules and excitonic fine structure.

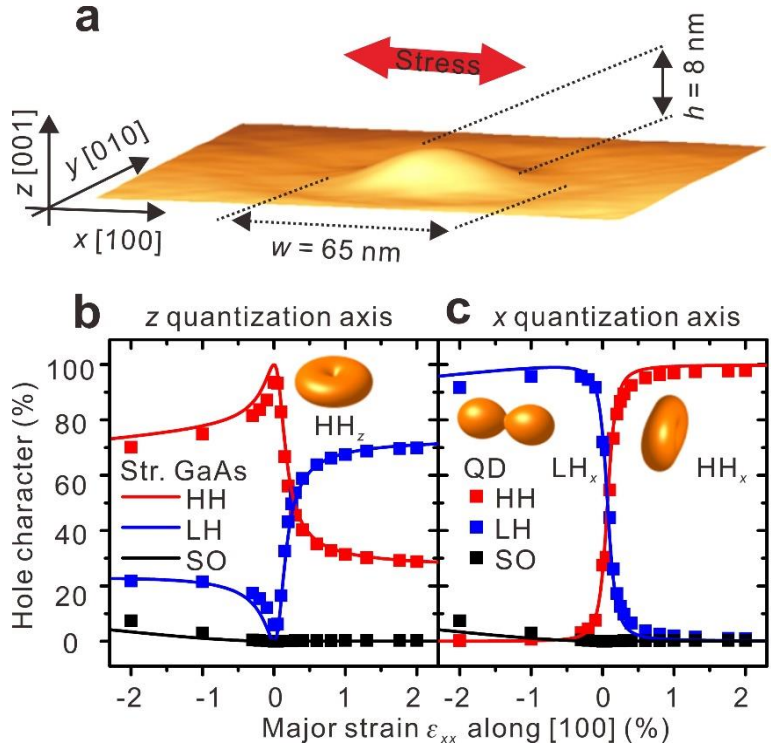

**Figure 1. Illustration of the concept used to rotate the natural quantization axis of an epitaxial QD. a**, 3D view of an AFM image of a GaAs QD embedded in AlGaAs matrix subject to uniaxial stress. **b, c**, Calculated effect of uniaxial stress on the degree of mixing of the topmost VB using the $z$ and $x$ quantization axis, for bulk GaAs subject to a fixed in-plane biaxial compression with $\sigma_{xx}=\sigma_{yy}=-120$ MPa (solid curves) and for the experimentally studied QDs (symbols). The plots illustrate the importance of the choice of the quantization axis when discussing VB mixing and show that the quantization axis of the chosen QDs can be flipped with moderate strains. Insets: Angular dependence of the probability density distribution of the Bloch wavefunctions of the topmost VB states at the $\Gamma$ point showing the conversion of a $HH_z$ state into a $HH_x$ ($LH_x$) state under tension (compression).

## Results

**Illustration of the concept**

As mentioned above, the HGS of epitaxial QDs has dominant $HH_z$ character as a consequence of the vertical confinement. The same situation is encountered when biaxial compression in the *x-y* plane is applied to bulk GaAs. In these cases, the angular dependence of the $HH_z$ Bloch wavefunction shows a "donut" shape (see inset in Fig. 1**b**). Combined with the s-like electron Bloch-wavefunction, such a state couples only to light with polarization perpendicular to the *z* quantization axis.

Obviously, for bulk GaAs, the quantization axis could be set to the [100] (*x*) direction by simply applying a biaxial stress in the *y-z* plane or uniaxial stress along the *x*-direction. Now the question is: is it possible to obtain a QD with an in-plane quantization axis through the application of realistic stress values? Previous experiments have shown that symmetry breaking in the *x-y* plane results in substantial $HH_z$ - $LH_z$ hole-mixing[27], but the possibility of reaching pure $HH_x$ or $LH_x$ states has not been discussed so far. And in fact the answer would be negative for conventional Stranski-Krastanow QDs because vertical confinement and in-plane compression (of the order of GPa) "team up" to stabilize a vertically-oriented quantization axis. Initially unstrained GaAs QDs with relatively low confinement energies are instead ideally suited to address the question. To quantify to what extent the quantization axis is oriented along the original *z* direction or the desired *x* direction under uniaxial stress, we calculated the HGS of our QDs (Fig. 1**a**) via the empirical pseudopotential method (EPM) and projected it onto the $HH_\mathbf{n}$, $LH_\mathbf{n}$, and $SO_\mathbf{n}$ (split-off) states, i.e. the eigenstates of the angular-momentum-projection operator $J_\mathbf{n}=\mathbf{J}\cdot\mathbf{n}$ along the quantization axis direction specified by the unit vector **n**. The results for **n** parallel to the *z*- (*x*-) directions are shown with symbols in Fig. 1**b** (1**c**). Using the conventional quantization axis *z* [Fig. 1**b**], we would reach the wrong conclusion that uniaxial stress results in strong HH-LH mixing even for large strains. By using instead the new quantization axis *x*, we see that the topmost VB has almost pure $HH_x$ character (with some $LH_x$ admixture) upon sufficiently strong tension and almost pure $LH_x$ character (with some $SO_x$ admixture) upon compression [Fig. 1**c**]. It is important to note that for intermediate values of strain the overlap of the HGS with the eigenstates of $J_\mathbf{n}$ is rather poor for any **n**, indicating that the HGS has low symmetry and a "good" quantization axis cannot be defined. This also means that uniaxial stress along *x* produces a quantization-axis flip rather than a smooth rotation. (For more details, see Supplementary Notes 8-10 and Supplementary Movie.) Remarkably our EPM calculations predict that the "swapping" of quantization axis occurs already at moderate strains: the HGS of our QDs should have >90% HH$_x$ character for strains $\varepsilon_{xx} \gtrsim 0.3\%$ and >90% LH$_x$ character for $\varepsilon_{xx} \lesssim -0.1\%$. This result crucially relies on the use of tall and initially unstrained QDs and could not be achieved with conventional Stranski-Krastanow QDs (confirmed by experiments not shown here). The predicted evolution of the HGS upon uniaxial stress is robust and can be even caught with a simple model of bulk GaAs subject to a fixed biaxial stress in the *x-y* plane and variable uniaxial stress along the *x* direction, see curves in Figs. 1**b, c**. Under tension we expect a donut-shaped Bloch wavefunction (right inset in Fig. 1**c**), a configuration suitable for light-coupling into *x*-oriented waveguides designed to sustain TE-like or TM-like modes (the HH$_x$–exciton emission is dominated by a *y*- and a *z*-oriented dipole). Under compression the wavefunction has instead a dumbbell shape elongated along the *x*-direction (left inset in Fig. 1**c**). This configuration is well suited for coupling into y-oriented waveguides designed to sustain TE-like modes (the LH$_x$–exciton emission is dominated by an *x*-oriented dipole).

Before discussing the future integration of stress-engineered QDs into quantum-photonic circuits we present below the experimental proof of the concept.

**GaAs QDs under uniaxial tensile stress** To test experimentally the above concept and to follow the evolution of the emission of single QDs under uniaxial stress, we have developed a piezoelectric actuator capable of "stretching" an overlying semiconductor layer up to its mechanical fracture. The actuator is made of [Pb(Mg$_{1/3}$Nb$_{2/3}$)O$_3$]$_{0.72}$[PbTiO$_3$]$_{0.28}$ (PMN-PT) and combines the advantages of a strain-amplifying suspension platform[28,29] and continuously-variable stress[26,30]. PMN-PT substrates were micro-machined to feature two fingers separated by a narrow gap, see Fig. 2**a**. A negative voltage $V_p$ applied to electrodes placed at the bottom of the two fingers leads to their in-plane contraction and consequent gap expansion. While the maximum strain achievable in PMN-PT is limited to about 0.2% at low temperatures[26], strain in a layer suspended between the fingers [here a 250-nm-thick (Al)GaAs membrane containing GaAs QDs] is amplified by a factor $2l/d$, where *l* is the length of each finger along the *x*-axis (here 1.5 mm) and *d* is the gap width (20~60 μm). Our simple actuator is thus capable of delivering strain values comparable to state-of-the-art microelectromechanical systems[31] and provides a compact alternative to the commonly used bending method[32–34] for operation at cryogenic temperatures [here ~8 K for all measurements] in a cold-finger cryostat.

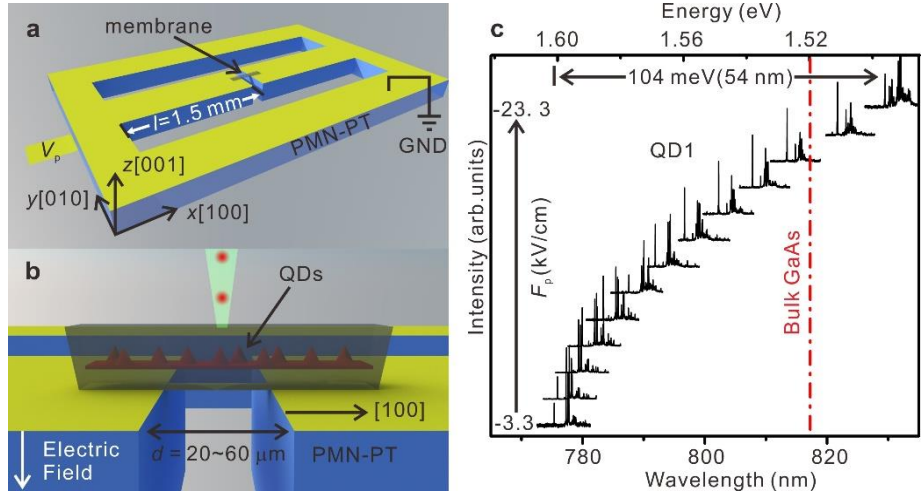

**Figure 2. Experimental configuration and emission energy tuning of a GaAs QD via uniaxial stress provided by a micro-machined PMN-PT actuator. a**, Sketch of the actuator featuring two fingers with length $l$ separated by a gap of width $d$. A semiconductor membrane with embedded QDs is bonded on the fingers and forms a bridge above the gap. $V_p$ is the voltage applied to the bottom of PMN-PT actuator with respect to the top contact, which is grounded. Because of the chosen poling direction, a negative electric field $F_p$ across the PMN-PT induces a contraction of the fingers and uniaxial tensile stress in the semiconductor along the $x$ direction. The coordinate system is the same as for the (Al)GaAs crystal. **b**, Side-view of the device. PL measurements are performed by exciting and collecting PL along the $z$ axis. **c**, Normalized PL spectra of a GaAs QD measured for increasing uniaxial tensile stress (from bottom to top). The large tuning range leads to emission below the bandgap of unstrained bulk GaAs (dashed line).

Strain amplification allows the emission energy of a QD to be continuously shifted in a spectral range exceeding 100 meV, as illustrated by the photoluminescence (PL) spectra of Fig. 2**c**. This shift, which is the largest achieved so far for QDs integrated on piezoelectric actuators, overcompensates the confinement energies and leads to emission well below the bandgap of bulk GaAs [see dashed line in Fig. 2**c**]. From the comparison between the observed shift and the one calculated with EPM and the configuration interaction (CI) method, we deduce a maximum stress of ~1.3 GPa (strain ~1.5%), a value which is limited by mechanical fracture of the membrane.

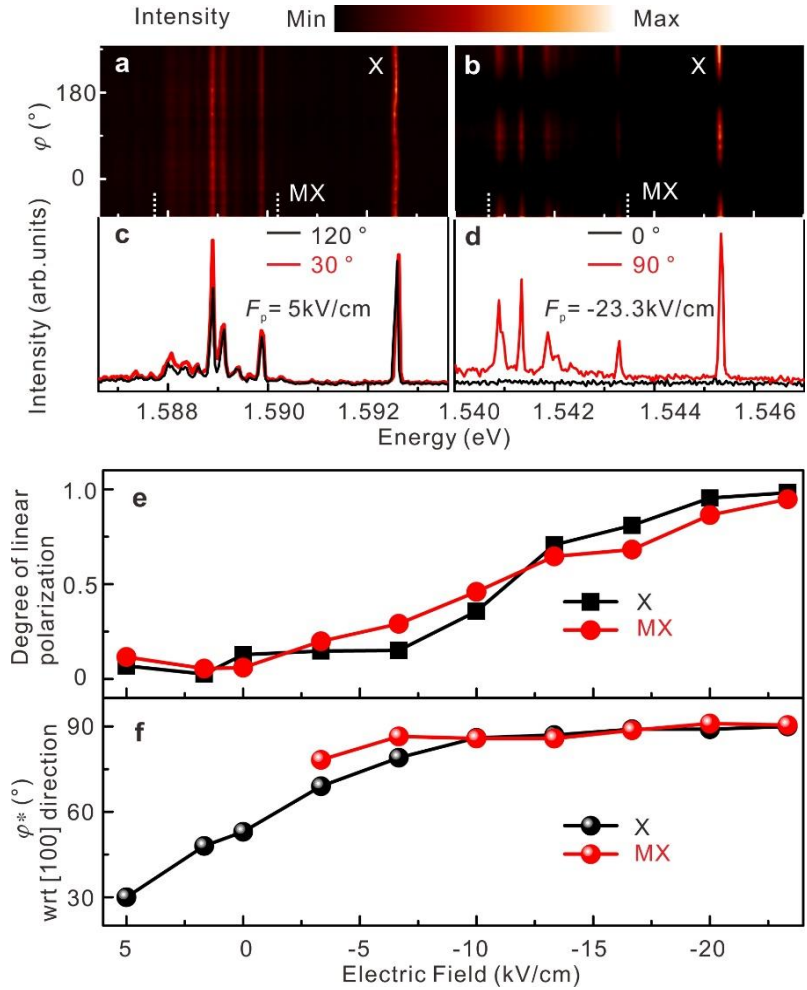

**Figure 3. Rotation of natural quantization axis under uniaxial stress along [100]. a-d**, Color-coded linear-polarization resolved PL spectra of a GaAs QD for increasing tensile stress (**a**, **b**) and selected spectra along orthogonal directions (**c**, **d**). The polarization angle $\varphi$ is referred to the [100] crystal direction. Initially, (**a**, **c**), the emission from the neutral exciton X and multiexcitons MX shows no significant net polarization and the X emission is characterized by two bright components linearly polarized along random directions (30° and 120° for this QD). This random orientation stems from slight anisotropy in the confinement potential defined by the QD and also from some process-induced prestress. At large stress, (**b**, **d**), the PL is almost fully polarized along the [010] direction ($\varphi^*=90°$). **e,** Evolution of the degree of linear polarization of X emission and MX "band" (emission between vertical dashed lines in **a** and **b**) for increasing tensile stress (increasing magnitude of applied electric field on actuator). **f,** Evolution of polarization orientation $\varphi^*$ for the MX band and of the high energy component of the X line for varying stress.

We now focus on the experimental proof that uniaxial tensile stress along the [100] (*x*) direction leads to a HGS with $HH_x$ character, manifesting in characteristic optical selection rules. To this aim, we performed linear-polarization-resolved measurements of the PL of various GaAs QDs under increasing stress while collecting light emitted along *z*, see sketch in Fig. 2. Figure 3**a** shows color-coded PL spectra of an almost unstrained QD embedded in a large (~2×4 mm$^2$) membrane. (A small positive electric field $F_p$ was applied to the actuator to partially compensate for some residual processing- and cooling-induced stress). The neutral exciton (X) emission shows a wavy pattern as a function of polarization direction $\varphi$ due to the "fine-structure-splitting"[35] (FSS) caused by slight in-plane anisotropy of the confinement potential[36,37]. Several multiexcitonic (MX) lines are also observed, which stem from recombination of a ground-state electron with a HGS (as for X), but in presence of additional photo-generated carriers in the QD. Overall, the initial spectra [Fig. 3**a, c**] shows no net polarization in the *x-y* plane, indicating weak $HH_z$−$LH_z$ mixing[27,38,39]. The scenario changes completely under strong tension, as shown in Fig. 3**b, d** for the same QD: Light from X and MX becomes fully polarized parallel to the y-direction. This is exactly what we expect for a $HH_x$−exciton, since the limited numerical aperture of the used objective (0.42) prevents us collecting *z*-polarized light. The degree of linear polarization, defined as $P = (I_{\max} - I_{\min})/(I_{\max} + I_{\min})$, with $I(\varphi)$ being the integrated intensity for the X or the MX lines, is shown in Fig. 3**e**. *P* increases with increasing stress and the angle $\varphi_*$ for which $I(\varphi_*)=I_{\max}$ aligns with the *y*-direction (perpendicular to the *x* quantization axis for the $HH_x$ state), see Fig. 3**f**. Because of the competition between intrinsic VB-coupling induced by the low symmetry of the nanostructure and the strain-induced effects, the transition from a $HH_z$ to a $HH_x$ is smooth, as already anticipated in Figs. 1**b**, **c**. However, for sufficiently large stress, the in-plane optical dipoles of X and MX transitions are deterministically aligned perpendicular to the pulling direction, independent of the initial orientation, as shown in Fig. 3**f** and Supplementary Notes 4 and 6.

The picture is not yet complete: an $HH_x$ - exciton couples not only to *y*-polarized but also to *z*-polarized light. To observe such a signature without resorting to side collection[27,40] and to study in detail the evolution of the X fine structure we have used narrow membranes (~3.5 µmwidth) with tilted edges (obtained by wet chemical etching, see Supplementary Note 5), which we expect to partially deflect *z*-polarized light into the collection path [see sketch in Fig. 4**a**]. After collecting and averaging polarization-resolved PL spectra of a QD for a fixed value of the voltage applied to the actuator, we shifted them horizontally using one of the bright exciton lines (indicated with $B_{y'} \rightarrow B_y$ in the following) as a reference. The result for a representative QD is shown in Fig. 4**b**. (We have measured in detail the behavior of 2 additional QDs finding fully consistent results, see Supplementary Note 6). Initially the emission is characterized by two lines ($B_{x',y'}$), which are linearly polarized perpendicular to each other in the *x-y* plane and are split by the FSS, see Fig. 4**c**. This is the typical signature of a $HH_z$ exciton. For the investigated QDs, the orientations *x'* and *y'* is rather random [see, e.g., Fig. 3**a**] and the average FSS is ~4 µeV before processing[20]. We stress again that fluctuations in the azimuthal orientation of the transition dipoles are commonly encountered also for other QDs[19] and represent a problem if light has to be efficiently coupled into a waveguide. With increasing stress, the energy separation between the two lines increases, the polarization direction of the high(low) energy component rotates and aligns to the *y*(*x*) direction independent of the initial orientation (see also Fig. 3**f**), and the intensity of the low energy component drops monotonically. The latter observation indicates that $B_{x'}$ converts to a dark state ($D_x$). We note that in general $B_{x',y'}$ undergo an anticrossing at moderate strain levels, as long as the stress orientation does not coincide with the anisotropy axis defined by the confinement[26]. However, different from the behavior in the perturbative regime, in which the "FSS" varies linearly with the applied stress[41] away from the anticrossing, the ($B_{y'} \rightarrow B_y$)-($B_{x'} \rightarrow D_x$) splitting almost saturates at a value of ~200 µeV for all investigated QDs.

The most striking feature emerging from Fig. 4**b** is an additional line appearing first as a shoulder at the low energy side of $B_{y'} \rightarrow B_y$ [$D_z$ in Fig. 4**d**] and then moving to its high energy side with increasing stress. Since such a component appears as *y*-polarized and was not observed in any QDs embedded in large membranes, we can safely attribute it to an initially dark exciton $D_z$, which evolves into a *z*-polarized bright exciton $B_z$. The final configuration at large stress [Fig. 4**e**] is fully consistent with a $HH_x$ HGS: two bright excitons ($B_y$ and $B_z$), which have polarization perpendicular to the new quantization axis *x* and two dark excitons. In the ideal case of pure $HH_x$ states, the Bloch wavefunctions of such dark states (one of which remains dark throughout the experiment) are a linear combination of the states with $J_x = \pm 2$. The large "FSS" ($B_y$ - $B_z$ splitting) can be qualitatively attributed to the strong anisotropy of the QD confinement potential in the *y-z* plane [see Fig. 1**a**], leading to a low(high) energy component $B_y(B_z)$ parallel to the long(short) axis of the QD, compatible with results obtained with $HH_z$-excitons confined in GaAs QDs with elongation in the *x-y* plane[37]. We note that the achieved configuration is markedly different from that of $LH_z$-excitons obtained either in vertically elongated InGaAs nanostructures[42] or in GaAs QDs under in-plane biaxial tension[23]. In that case the quantization axis is still along the *z*-direction and there are three optical dipoles (instead of two for the $HH_x$ case): one is aligned along *z* and the other two lie, randomly oriented, in the *x-y* plane.

Figure 4**g** shows the evolution of the X emission under tensile stress along [100] as calculated with EPM+CI. As in the experiment, energies are referred to the $B_y$ line. The size of the symbols is proportional to the strength of each transition. The calculations qualitatively reproduce all the experimentally observed features including the darkening of one of the initially bright excitons and the brightening of one of the initially dark excitons. Also the energy shifts follow the experiment, although the absolute magnitudes are systematically smaller - a discrepancy, which is currently under investigation. Finally we note that the rotation of the quantization axis and the associated donut-shaped HH states from the *z* to the *x* direction is accomplished through a complex topological transformation of the HGS wavefunction, as reflected by the angular dependence of the electron-HGS transition-dipole calculated by EPM+CI, see Fig. 4**f**. [Because the electron wavefunction is marginally affected by stress, such plots closely resemble the angular dependence of the Bloch wavefunctions shown in Fig. 1**c** and in the Supplementary Movie].

## Discussion

We conclude by discussing the possible implications of this work for integrated quantum photonics[4,12]. Particularly relevant features of quantum emitters for such applications are: (i) efficient light-coupling to guided modes; (ii) fast radiative recombination, which is beneficial for high photon rates and indistinguishability. The re-orientation of the quantization axis of a GaAs QD shown here addresses the first point. We stress here that the used QDs have excellent optical properties in terms of single-photon purity and indistinguishability when excited resonantly through a 2-photon-absorption process[24,43,44], which is compatible with planar photonic circuits and strongly reduces exciton recapture[44,45]. The fact that no line broadening was observed up to the maximum achievable stress values [see, e.g. Fig. 4**c-e**] makes us confident that stress along [100] does not deteriorate such excellent properties.

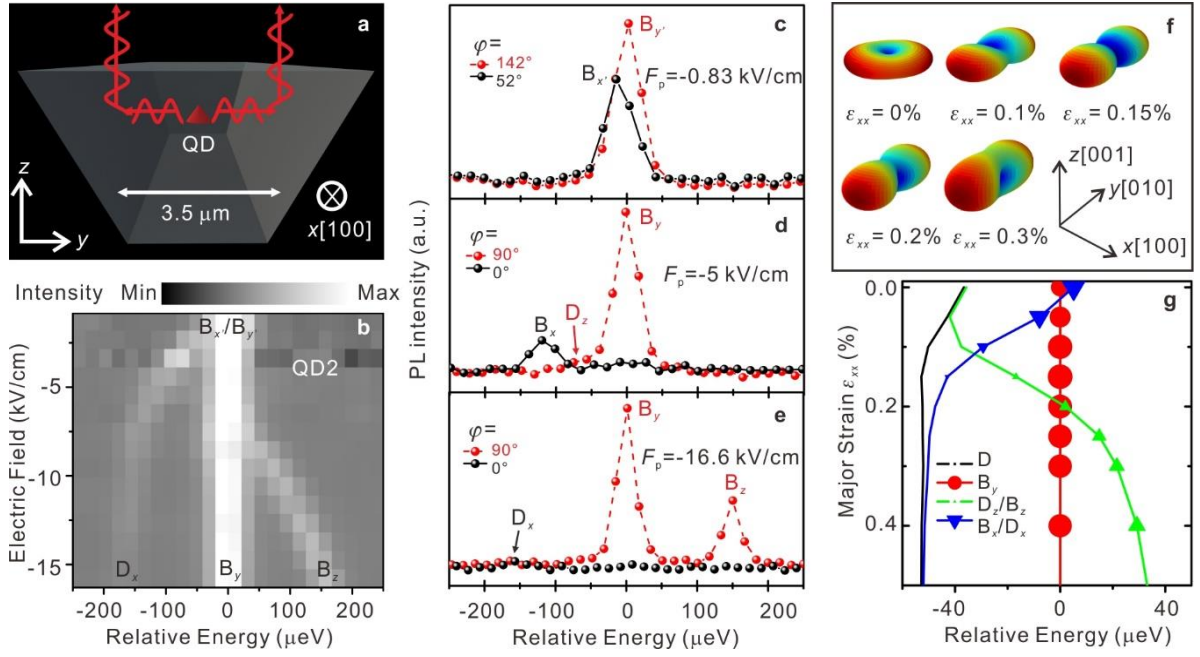

**Figure 4. Evolution of the fine structure of a neutral exciton confined in a GaAs QD for increasing uniaxial stress and comparison with theory. a**, Sketch of the cross-section ($y$-$z$ plane) of the stripe-like membrane used to project the vertically polarized component $B_z$ into the collection optics [see Fig. 2**b**], the pulling ($x$) direction is orthogonal to the $y$-$z$ plane. From the geometry, we expect such light to appear as $y$-polarized. **b,** PL spectra of the neutral exciton (X) emission of a QD as a function of electric field applied to the piezoelectric actuator. Spectra are shifted horizontally using the line labeled as $B_{y'}$ or $B_y$ as reference. **c-e,** Representative linearly polarized PL spectra of the X emission in a QD for different values of the field $F_p$ applied to the actuator (uniaxial stress). In **c** we see the usual fine-structure-splitting (FSS) of a $HH_z$ exciton, characterized by two orthogonally polarized lines. The observed polarization anisotropy, relatively large FSS (14 µeV), and random orientation of the polarization directions are ascribed to $HH_z$-$LH_z$ mixing induced by QD anisotropy and process-induced prestress. The applied uniaxial stress aligns the transition dipoles along the crystallographic axes, as shown in **d** and **e**. $z$-polarized emission appears first as a shoulder on the low energy side of the $B_y$ line (marked as $D_z$ in **d**). The emission of a $HH_x$ exciton is shown in **e**. **f**, Polar-coordinate representation of the excitonic transition dipole calculated by the EPM+CI. **g**, Relative transitions energies of the four excitonic components under uniaxial stress (compressive, top

and tensile, bottom) computed by EPM+CI. The symbol size is proportional to the oscillator strength of the transitions.

Besides the possibility of orienting the transition dipoles perpendicular to the propagation axis, there are two additional appealing features of uniaxial stress engineering. First, under proper uniaxial tension (which may be achieved by suspending the QD source at the edge of a photonic chip followed by coating with a dielectric stressor[46]) it may be possible to achieve level degeneracy between the $B_y$ and $B_z$ lines and obtain a QD launching polarization entangled photon pairs in a photonic circuit using the biexciton - $HH_x$ exciton cascade. Second, if we inspect the calculated transition strengths of excitonic dipoles under uniaxial compression (shown in Fig. 5a but not yet explored here experimentally), we see that the emission of a $LH_x$-exciton is dominated by a single emission line with polarization parallel to the compressive stress direction and with an oscillator strength which is almost twice that of each of the two bright excitons in our $HH_z$ QDs. Considering that $HH_z$ excitons in GaAs QDs are already characterized by very short lifetimes (~250 ps)[24,47], the combination of deterministic QD positioning in photonic structures[48,49] and uniaxial-strain-engineering (see sketch in Fig. 5b) may lead to ideal single-photon sources with enhanced recombination rate [~8 GHz, see right axis of Fig. 5a] and indistinguishability levels. Instead of resorting to the Purcell effect[7,11] lifetime reduction would rely here on the strain-induced suppression of $y$ and $z$-polarized emission and consequent "oscillator-strength concentration" on the $x$-polarized emission. Even if technologically challenging, replacing the static stressors shown in Fig. 5b with piezoelectric materials would allow fine tuning the emission of different QDs to the same energy and to achieve scalable photon sources for integrated quantum photonics circuits.

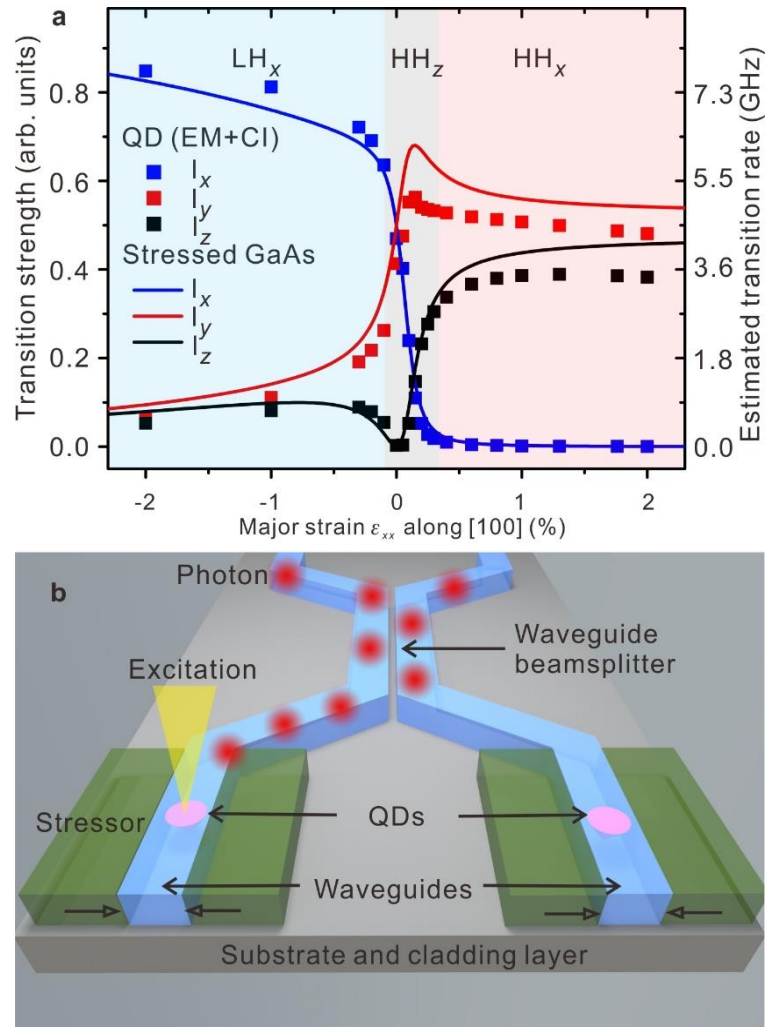

**Figure 5. Envisioned applications of strain-engineered single photon source for integrated quantum photonic circuits. a**, Calculated transition strength and estimated rates for light polarization along the *x* (blue), *y* (red), and *z* (black) directions for varying uniaxial stress along *x* for QDs with structure taken from experiment (symbols) and GaAs subject to a fixed biaxial compression with $\sigma_{xx}=\sigma_{yy}=-120$ MPa (solid lines). **b**, Envisioned approach to obtain high-speed single-photon sources based on LH$_x$ excitons confined in GaAs QDs. A waveguide is fabricated to contain preselected QDs followed by deposition of side dielectric layers, which act at the same time as cladding, passivation and stressor layers. The uniaxial stress (produced, e.g. by the different thermal expansion coefficients of dielectric and (Al)GaAs heterostructure aligns the quantization axis of the QD along the stress direction (*x*) and the resulting LH$_x$ exciton acts as an ultrafast source of single photons ideally matched to the propagating modes in the waveguide.

## Methods

**Computational Methods**

For 3D calculations we used the quantum dot shape as directly provided by the digital form of a representative AFM image. We first relaxed the 6.5 million atomic positions, including the quantum dot and a sufficiently large $Al_{0.4}Ga_{0.6}As$ alloy barrier using the valence force field approach[50] to minimize the strain energy. The determined positions are used as input into the atomistic empirical pseudopotential method using the strained linear combination of bulk bands[51]. The ensuing single particle wavefunctions and eigenenergies are subsequently used into a screened configuration interaction framework[52] where fully correlated exciton and multiexciton states are calculated. The required Coulomb and exchange integrals are microscopically screened according to the bulk model of Thomas and Fermi. The dipole moments are calculated based on the correlated excitonic wavefunctions[53]. A similar approach was used to compute the properties of light-hole ($LH_z$) excitons in Ref.[23] but with unscreened Coulomb and exchange integrals.

The curves shown in Figs. 1**b-c** and 5**a** were obtained from the Luttinger-Kohn and Pikus-Bir (PB) Hamiltonians including the HH, LH and the SO bands[54] by using the eigenstates of the PB Hamiltonian (at the Γ point) corresponding to the topmost twofold degenerate VBs. The effect of vertical confinement was introduced phenomenologically by adding a fixed biaxial compression. The strain values displayed in the abscissa of Figs. 1**b-c** and 5**a** are those corresponding to a variable uniaxial stress ranging from -2 to 2 GPa along [100]. Additional details and calculations are presented in the Supplementary Notes 8-12.

**Sample Growth**

The used sample was grown by molecular epitaxy (MBE) on semi-insulating GaAs (001) substrate and consists of GaAs QDs placed at the center of a symmetric slab consisting of 30 nm $Al_{0.4}Ga_{0.6}As$, 90 nm $Al_{0.2}Ga_{0.8}As$ and 5 nm GaAs on top and below the QDs. This active structure was grown on a 100-nm-thick $Al_{0.75}Ga_{0.25}As$ etch-stop/sacrificial layer, which was removed via selective etching for membrane transfer. On the bottom $Al_{0.4}Ga_{0.6}As$ barrier, nanoholes were produced by the Al-droplet-etching method. During subsequent deposition of 1.6 nm GaAs and annealing, such nanoholes act as a mold cast to obtain GaAs QDs. For additional details, see Supplementary Note 1.

**Device fabrication**

300-µm thick, (001)-oriented, PMN-PT substrates were cut by a micromachining system equipped with a femtosecond laser[55,56]. After cleaning, the top surface of the resulting actuator was homogeneously coated with a Cr/Au layer acting as top (electrically grounded) electrode. On the bottom surface only the two PMN-PT fingers were metallized leaving the surrounding frame uncoated, so that electric-field-induced deformation is limited to the fingers. In order to integrate the semiconductor structures on the micromachined actuators we first used photolithography and wet chemical etching to define mesa structures with the desired lateral size and shape (in our experiment we used two structures, one is about ~2×4 mm$^2$ and another is 3.5×300 µm$^2$). The sides of such mesas are tilted because of photoresist undercut during etching, see Supplementary Note 4. After lithography, the [100] crystal direction of the (Al)GaAs heterostructure was aligned parallel to the actuator fingers. Bonding of such layers onto the actuator was achieved via a flip-chip process using SU8 photoresist as adhesive bonding layer because of its high bond strength and low process temperature. A series of chemical etching steps were then performed to remove the original GaAs substrate and $Al_{0.75}Ga_{0.25}As$ layer, leaving membranes bonded onto the PMN-PT actuator. We note that while the actuator shows a fully predictable behavior in the tensile regime, poorly-controlled membrane buckling prevents us to obtain reliable results in the compressive regime. For additional details, see Supplementary Notes 2-3.

**Optical Characterization**

For optical measurements, the device was mounted on the cold-finger of a He-flow cryostat and cooled to ~8 K. A continuous wave laser (with wavelength of 532 nm) was used to achieve above-bandgap excitation of carriers. A 50× microscope objective with 0.42 numerical aperture was used on top of the sample to both focus the laser and collect the PL signal. After passing through a half-wave plate and a fixed linear polarizer, the PL signal was analyzed with a 750-mm focal length spectrometer equipped with a 1800 lines/mm ruled grating and a liquid-nitrogen cooled Si CCD with 20 µm pixel size. Piezo creep was observed during the measurements, especially at high electric fields. Although active stabilization could be used,[26] the effect on polarization-resolved spectra was compensated here after data acquisition, as discussed in the Supplementary Note 7. The reproducibility of the presented results was thoroughly tested by repeating similar experiments of several QDs, see Supplementary Notes 4 and 6.

**Data Availability**

The data that support the findings of this work are available from the corresponding author upon request.

**Acknowledgements**

This work was supported by the FWF (P 29603), the Linz Institute of Technology, the BMBF (Q.com-H, Contract No. 16KIS0108), the EU project HANAS (No. 601126210), AWS Austria Wirtschaftsservice (PRIZE Programme, Grant No. P1308457) and the European Research Council (ERC) under the European Union's Horizon 2020 Research and Innovation Programme (SPQRel, Grant agreement No. 679183). X. Yuan was supported by China Scholarship Council (CSC, No. 201306090010). V. Křápek was supported by MEYS CR, Project (No. LQ1601). The authors thank J. Claudon (CEA, Grenoble), S. Portalupi (University Stuttgart), and M.A. Dupertuis (EPFL, Lausanne) for fruitful discussions, M. Reindl, D. Huber, J. Wildmann for help with the processing and optical characterization and A. Halilovic, A. Schwarz, S. Bräuer, U. Kainz, and E. Vorhauer for technical support.

**Author contributions**

XY fabricated the devices with the help of HH, CS, and JMS, and performed optical measurements with the help of RT. GP processed the piezoelectric actuators with support of JE. JMS tested the actuator concept. YH performed preparatory work and grew the sample, with support of AR and OGS. AR performed **k·p** calculations and interpreted the results with help of FWB, VK and GB. FWB performed atomistic pseudopotential calculations with support of GB. AR and XY wrote the manuscript with input from all the authors. AR conceived and coordinated the project.

**Competing Interests**

The Authors declare no competing interests.

**Quantization axis flipping of quantum dot for integrated quantum photonics**

**Supplementary Information**

**Yuan** *et al*.

**Supplementary Note 1. Sample Growth and Structure**

The sample was grown by solid-source molecular epitaxy (MBE) on semi-insulating GaAs (001) substrate. After oxide desorption, a 200 nm thick GaAs buffer was grown at a substrate temperature $T_{sub}$ = 580 °C. Then the substrate temperature was increased to 600 °C and a 100 nm thick $Al_{0.75}Ga_{0.25}As$ sacrificial layer was deposited, followed by 5 nm GaAs, which was used to smoothen the surface and avoid oxidation of the membranes after releasing them from the substrate. Subsequently, 90 nm $Al_{0.2}Ga_{0.8}As$ and 30 nm $Al_{0.4}Ga_{0.6}As$ layers were deposited in sequence. Nanoholes were then etched on the surface of this layer by droplet etching [1,2]: After stopping the arsenic supply, 0.5 monolayer Al were deposited so that the Al droplet can etch holes into the surface of $Al_{0.4}Ga_{0.6}As$ layer during a subsequent annealing step under As flux. To infill the nanoholes, we deposited 1.6 nm GaAs followed by 2 minutes annealing. The GaAs QDs formed in the nanoholes due to the diffusion of GaAs driven by capillarity effects [3]. Since the surface is practically flat after this step [4], the shape of the QDs can be assumed to coincide with that of the nanoholes. An inverted AFM image of such a nanohole is shown in Fig. 1(a) of the main text. The QDs were capped with 30 nm $Al_{0.4}Ga_{0.6}As$ layer, followed by 90 nm $Al_{0.2}Ga_{0.8}As$ layer. Finally, a 5 nm GaAs completes the growth. The whole structure of the as-grown sample is shown in Supplementary Figure 1.

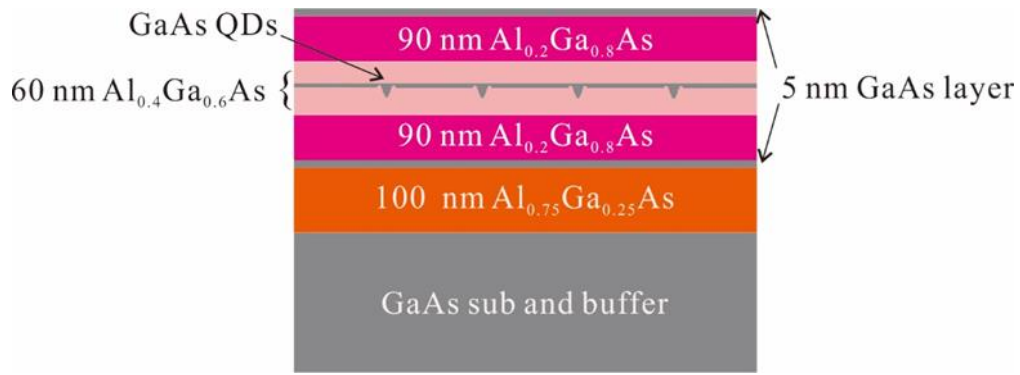

**Supplementary Figure 1**. Structure of the GaAs QDs sample used in this work.

**Supplementary Note 2. Device Processing**

500-µm thick, (001)-oriented, PMN-PT substrates (TRS Techonogies Inc.) were lapped and polished to a thickness of ~300 μm and cut by a commercial 3D-Micromac micromachining system equipped with a femtosecond laser[5,6]. The laser spot was focused down to 5 μm.

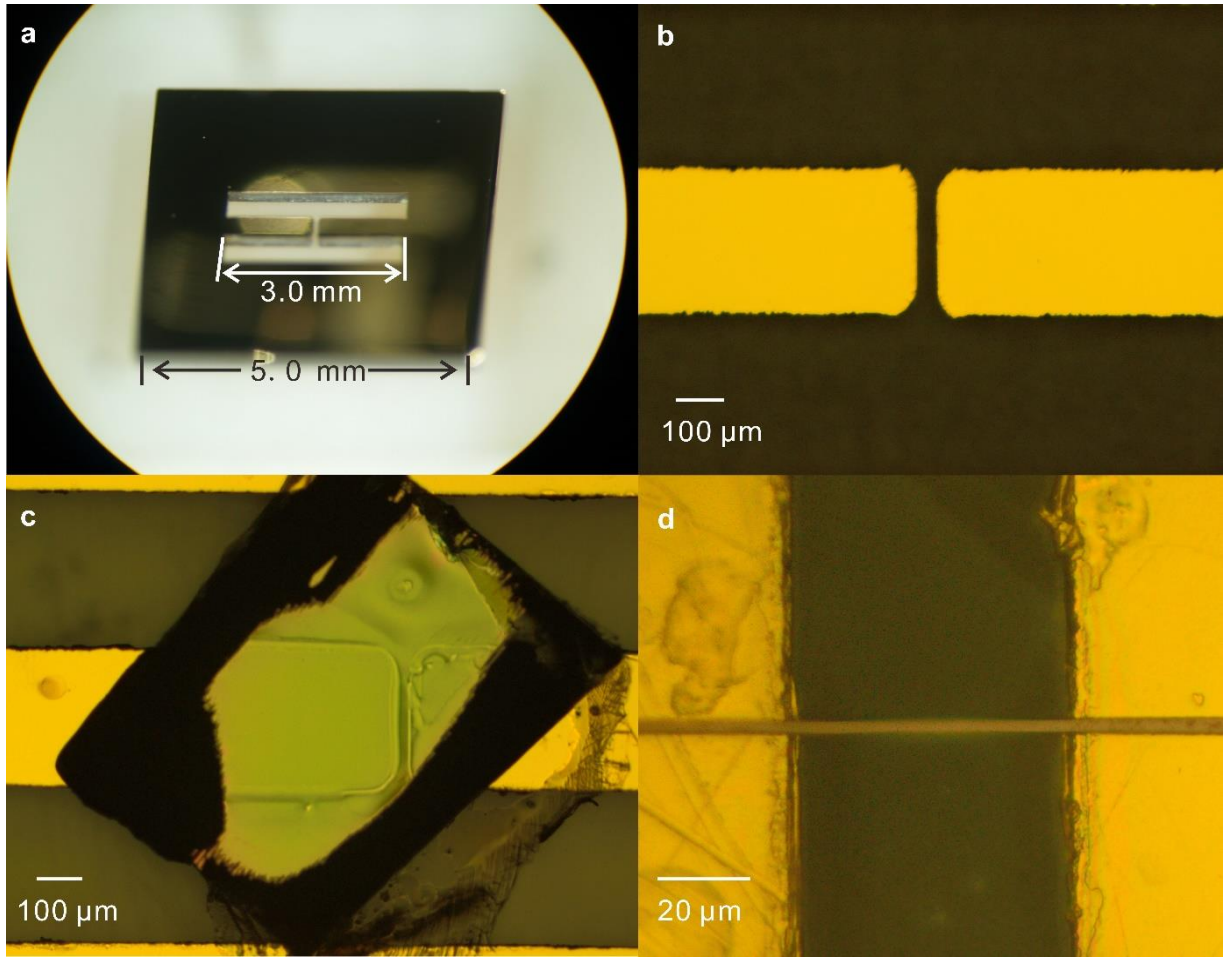

**Supplementary Figure 2**(a), Microscope image of the micro-machined "two fingers" PMN-PT actuator. (b), Enlarged microscope picture of the gap between the two fingers. (c), Micrograph of one device studied in this work, a large QDs-containing membrane bonded on the PMN-PT actuator (d), Microscope picture of another device featuring a narrow stripe membrane (3.5 μm in width ) bonded on the PMN-PT actuator.

A representative picture of a micro-machined "two-fingers" $[Pb(Mg_{1/3}Nb_{2/3})O_3]_{0.72}[PbTiO_3]_{0.28}$ (PMN-PT) actuator employed in our study is shown in Supplementary Figure 2(a). Because the total length of the two fingers (3.0 mm in total) is much larger than the length of the gap (20 ~ 60 μm) the relative size change of the gap is much larger than the relative deformation of the fingers when an electric field $F_p$ is applied across the fingers. This is the key why this novel micro-machined PMN-PT can provide large strain amplification. Supplementary Figure 2(b) shows a microscope image of the gap between the two fingers.

Information on the membrane processing and integration on the actuator can be found in Supplementary Note 5 and in Refs.[5,6]. Supplementary Figure 2(c) and (d) show the two different sizes of membranes which we studied in this work. Supplementary Figure 2(c) presents an optical microscope picture of the device with a large membrane used for the data shown in in Figure 3 in the main text and Supplementary Figure 2(d) shows an image of a device with narrow stripe membranes, used in Figure 2 and Figure 4 in the main text.

**Supplementary Note 3. Stress configuration and finite element simulations**

Since in the experiment we were not able to provide an independent measurement of the strain configuration produced by the piezoelectric actuator, we have performed finite element method (FEM) simulations. We provide below the relation between stress and strain in the case of ideal uniaxial stress along the [100] crystal direction of GaAs and the results of the numerical calculations.

Using the Voigt notation for the stress and strain tensors, the Hook's law for a cubic semiconductor reads:

$$\begin{pmatrix} \sigma_{xx} \\ \sigma_{yy} \\ \sigma_{zz} \\ \sigma_{yz} \\ \sigma_{xz} \\ \sigma_{xy} \end{pmatrix} = \begin{pmatrix} C_{11} & C_{12} & C_{12} & 0 & 0 & 0 \\ C_{12} & C_{11} & C_{12} & 0 & 0 & 0 \\ C_{12} & C_{12} & C_{11} & 0 & 0 & 0 \\ 0 & 0 & 0 & C_{44} & 0 & 0 \\ 0 & 0 & 0 & 0 & C_{44} & 0 \\ 0 & 0 & 0 & 0 & 0 & C_{44} \end{pmatrix} \begin{pmatrix} \varepsilon_{xx} \\ \varepsilon_{yy} \\ \varepsilon_{zz} \\ 2\varepsilon_{yz} \\ 2\varepsilon_{xz} \\ 2\varepsilon_{xy} \end{pmatrix} \quad (1)$$

Here the $\sigma_{ij}$ are the independent components of the stress tensor, $C_{ij}$ the components of the elastic stiffness tensor and $\varepsilon_{ij}$ the strain tensor. Since in our case the uniaxial stress is applied along the [100] direction, there is no shear stress, so that the above equation can be further simplified as:

$$\begin{pmatrix} \sigma_{xx} \\ 0 \\ 0 \end{pmatrix} = \begin{pmatrix} C_{11} & C_{12} & C_{12} \\ C_{12} & C_{11} & C_{12} \\ C_{12} & C_{12} & C_{11} \end{pmatrix} \begin{pmatrix} \varepsilon_{xx} \\ \varepsilon_{yy} \\ \varepsilon_{zz} \end{pmatrix} \quad (2)$$

$$\varepsilon_{yy} = \varepsilon_{zz} = -\frac{C_{12}}{C_{11}+C_{12}} \varepsilon_{xx} = -\nu_{[100]} \varepsilon_{xx} \quad (3)$$

Here $\nu_{[100]}$ is the Poisson ratio for uniaxial stress along the [100] direction ($\nu = 0.31$ for GaAs at a temperature of about 10 K[7]).

We used the commercial software package COMSOL Multiphysics to perform Finite Element Method (FEM) simulations for our actuators. All calculations assume linear elastic deformations. All the parameters are set based on one of the devices. We should mention that in the simulation we used the elastic and piezoelectric parameters of PMN-PT at room temperature. Since it is known that the piezoelectric constants of PMN-PT will decrease at cryogenic temperature[8], we adjusted the electric field applied to the PMN-PT to obtain strain magnitudes similar to our experiment.

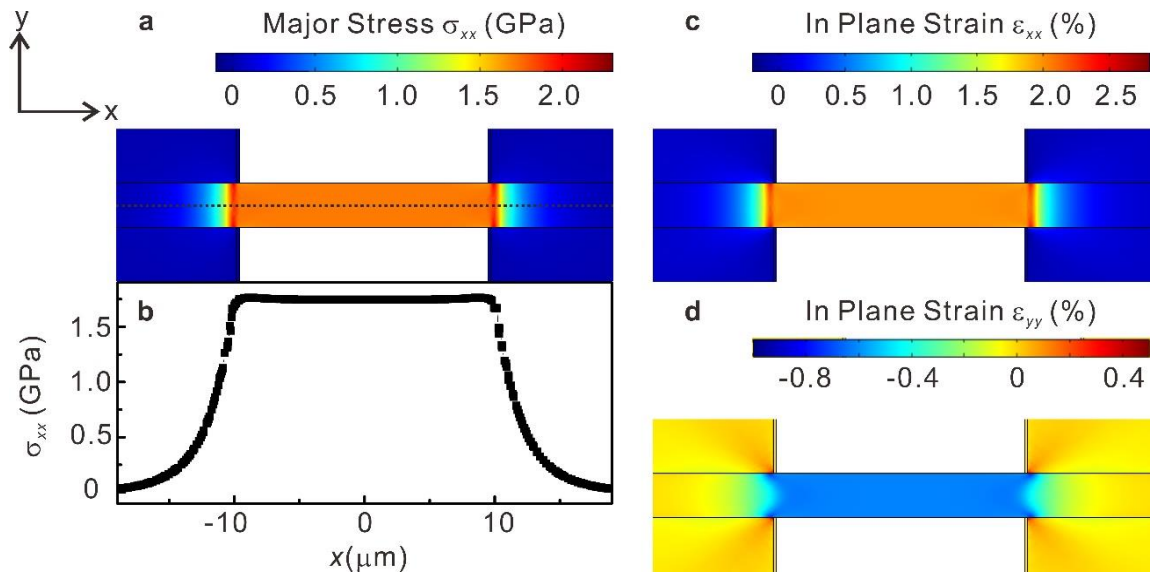

**Supplementary Figure 3**. FEM simulations of the stress and strain distribution in our devices. (a) Major stress $\sigma_{xx}$ map in the middle of the membrane in $z$ direction (the same location of the QDs). The dashed line shows the position of the line scan in (b). (c, d) In plane strain ($\varepsilon_{xx}$, $\varepsilon_{yy}$) maps, respectively. We verified that $\varepsilon_{zz}=\varepsilon_{yy}$, consistent with a uniaxial stress configuration.

Supplementary Figure 3(a) shows the major stress $\sigma_{xx}$ map in our device. It shows quite homogeneous tensile stress in the suspended area of the membrane. In order to have a clearer picture of the stress distribution along the $x$ direction of the membrane, we make a line scan [the line is locate in the middle of membrane, marked as dashed line in Supplementary Figure 3(a)] along $x$, see Supplementary Figure 3(b). The suspend area (-10 μm < $x$ < 10 μm) shows homogeneous tensile stress, while the tensile stress falls rapidly when we move to the area bonded on the PMN-PT actuator and changes into compressive stress when moving away from the gap (not shown here). As mentioned above, for an ideal uniaxial stress, $\varepsilon_{yy} = \varepsilon_{zz} = -\nu_{[100]}\varepsilon_{xx}$. Supplementary Figure 3(c) and (d) show the in plane strain maps of the strain tensor components $\varepsilon_{xx}$ and $\varepsilon_{yy}$ from the simulation. It is clear that in the suspend area there is tensile strain along *x* direction ($\varepsilon_{xx}$ = 2.0% > 0) and compressive strain along *y* direction ($\varepsilon_{yy}$ = -0.62% < 0), the Poisson ratio $\nu_{[100]}$ is about 0.31, consist with the expected value.

The impact of stress on the electronic properties of our QDs and bulk GaAs is described in the main text and also in Supplementary Note 9-12.

**Supplementary Note 4. Additional photoluminescence data on macroscopic membranes**

To confirm the reliability and reproducibility of the results shown in the manuscript (Fig. 3), measurements on another device with a large membrane were performed. The data collected on a randomly chosen QD located in the area of the membrane suspended above the actuator gap are shown in Supplementary Figure 4 Supplementary Figure 4(a) shows the color-coded PL spectra of the selected QD as a function of linear-polarization direction with no applied voltage to the actuator ($F_p$ = 0 kV/cm). In spite of the fact that no stress is intentionally applied to the membrane, we see slightly polarized emission from the neutral exciton (X) and multiexcitonic lines (MX) and a substantially larger X fine-structure-splitting (FSS) compared to the values obtained on as-grown samples. We attribute these two observations to residual (anisotropic, see below) stress arising from the device processing and from the different thermal expansion coefficients of the materials present in the device (PMN-PT, semiconductor, SU8 etc.), as mentioned in the main text. Under strong tension, see Supplementary Figure 4(b), all lines are nearly 100% polarized, similar to the results shown in Fig. 3(c) of the main text. In Supplementary Figure 4(c), we plot the degree of linear polarization *P* of X and MX as a function of $F_p$ (tensile stress). For the MX lines we integrated the intensity of all lines. It is obvious that *P* increases with the increasing tensile stress, which is consistent with the experimental and theoretical analysis in the main text.

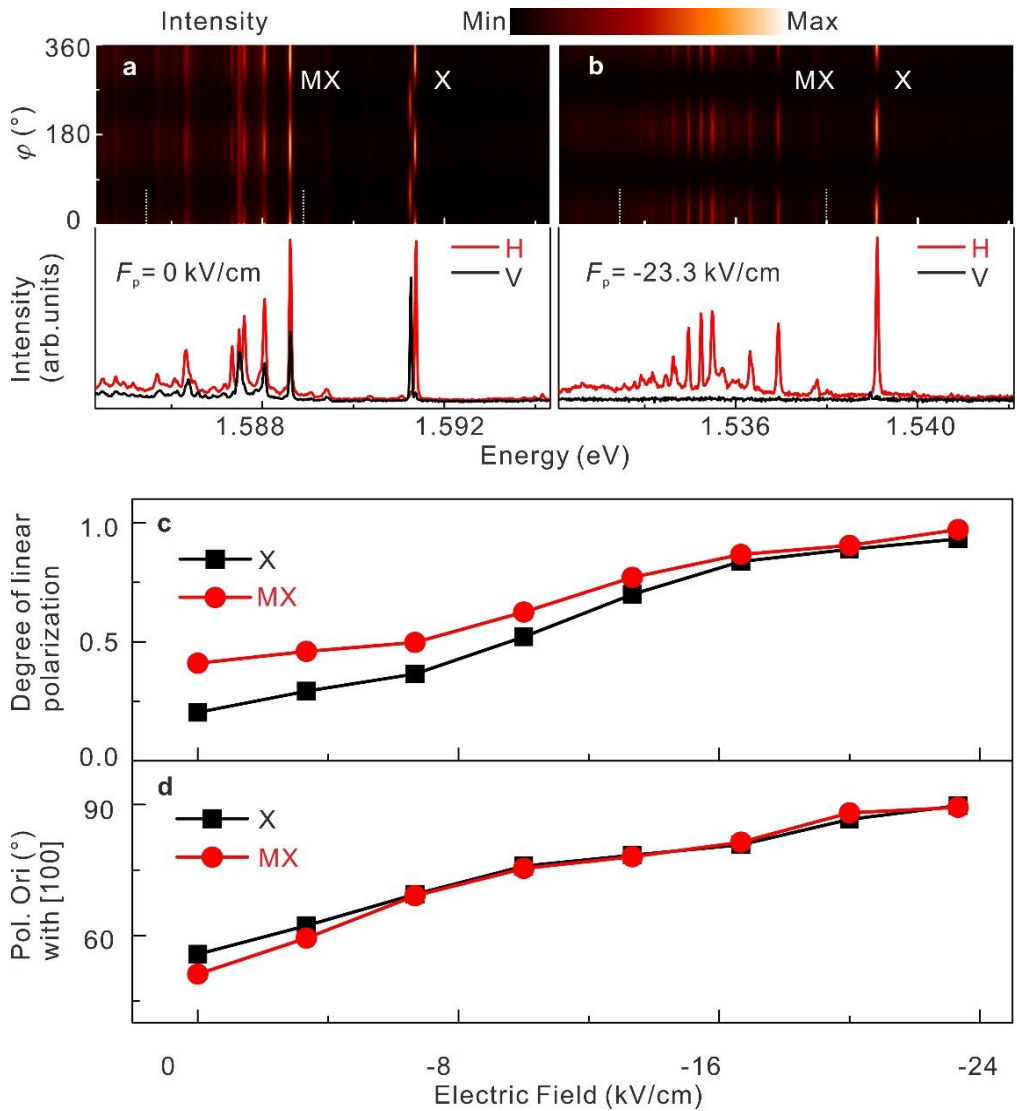

**Supplementary Figure 4**. (a-b) Color coded linear-polarization-resolved PL spectra of a GaAs QD at 0 applied field across the actuator and under high tensile stress. X and MX stand for neutral exciton and multiexcitonic lines. The slight polarization anisotropy observed in (a) is attributed to process- and cooling- induced stress. (c) Evolution of polarization degree with increasing tensile stress (magnitude of electric field $F_p$). (d) Evolution of the polarization orientation of X and MX for increasing $|F_p|$. While the initial polarization is randomly oriented, light becomes fully polarized parallel to the *y* direction (perpendicularly to the pulling direction) at large stress.

The presence anisotropic prestress combined with random fluctuations in the confinement potential defined by the QD lead to emission which is partially linearly polarized along a random direction. In addition, the transition dipoles for the X emission have random orientation even in absence of prestress[2,4]. In Supplementary Figure 4(d) we plot the polarization direction $\varphi^*$ along which the emission intensity is maximal as a function of the electric field across the actuator fingers (which we expect to be approximately proportional to the applied uniaxial stress). Independent on the initial polarization direction and the initial orientation of the X-dipoles, the emission lines (X and MX) become fully polarized perpendicular to the pulling direction defined by the actuator at large uniaxial stress, which is consistent with the theoretical prediction in main text.

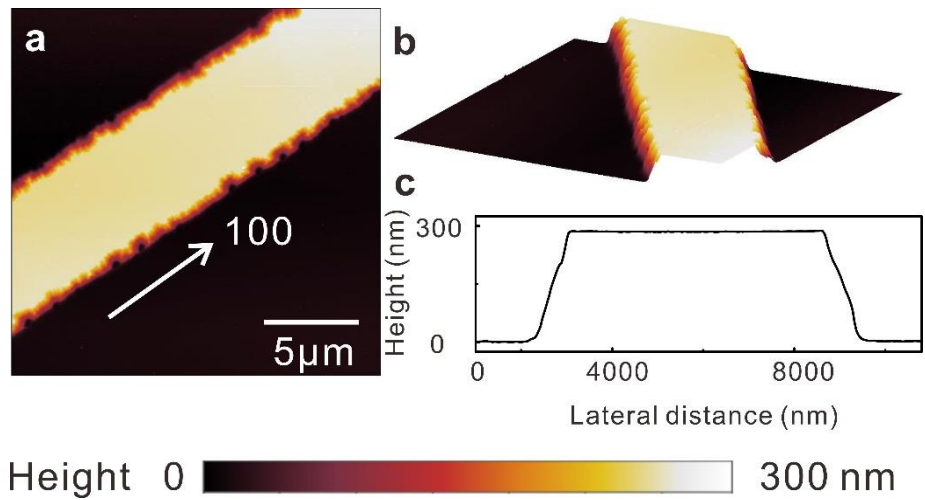

**Supplementary Figure 5**. (a) AFM image of stripe-like membrane after $H_2SO_4$:$H_2O_2$:$H_2O$ etching (b) 3D view of the AFM image of etched membrane. (c) Line profile of etched membrane.

**Supplementary Note 5. AFM topographic image of stripe membranes**

The etching solution used to produce mesa structures on the semiconductor sample consists of $H_2SO_4$:$H_2O_2$:$H_2O$ (1:8:200 in volume ratios). Etching was performed at room temperature (RT). This etching solution is widely used for GaAs[9] and is slightly anisotropic. During etching, the photoresist is undercut, leading to tilted side-walls, as illustrated by the AFM image in Supplementary Figure 5(a) and (b) and corresponding line scan in Supplementary Figure 5(c). The membrane is bonded upside-down on the actuator, so that the tilted edges act as rough reflectors, which allow us to collect *z*-polarized light from the sample top. Because of the poorly defined geometry and roughness, a quantitative evaluation of the relative intensity of *y*- and *z*-polarized components is not possible using this simple strategy.

However we expect *z*-polarized light to be "converted" in *y*-polarized light when detected from the top of the stripe, consistent with the results shown in Supplementary Figure 6 and Fig. 4 of the main text.

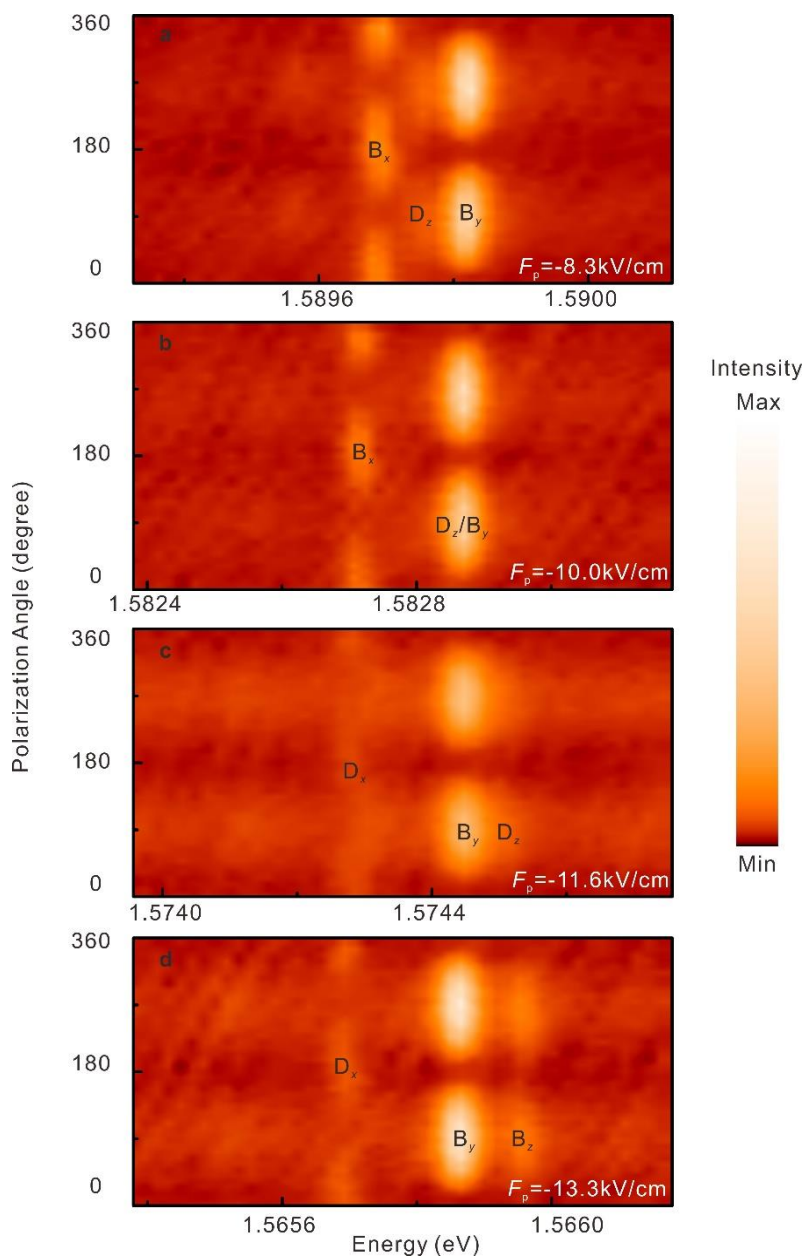

**Supplementary Figure 6**. Color-coded polarization-resolved PL signal of a neutral exciton emission in a GaAs QD for different values of the electric field applied to the actuator. Tensile stress increases from top to bottom. An angle of 0° corresponds to polarization along the pulling direction *x*. $D_z$ stands for one of the initially dark states, while $D_x$ is the dark exciton state under tensile stress, $B_y$ and $B_z$ are the new two bright exciton states under tensile stress. The chosen color-scale enhances also weak features.

**Supplementary Note 6. Linear-polarization-resolved PL spectra of neutral excitons confined in QDs embedded in stripe membranes**

Supplementary Figure 6 shows a set of polarization-resolved PL spectra of a neutral exciton X confined in a GaAs QD (QD1) under different tensile stresses [the same dot shown in Figure 2(c) of the main text]. These spectra provide a clear picture of the evolution of the X fine structure during the transition from a $HH_z$ to a $HH_x$ hole ground state (HGS). With increasing uniaxial stress, the initially dark exciton $D_z$ becomes bright [Supplementary Figure 6(a)] and gradually moves from the low energy side of the bright exciton $B_y$ to its high energy side, and finally becomes a bright exciton $B_z$ [Figure S6(d)]. The $z$-polarized component ($D_z$ or $B_z$) is reflected by the sidewalls of stripe membrane and appears to have polarization approximately parallel to the $y$ direction, as expected. The initially bright exciton $B_x$ moves to the low side energy side of $B_y$ with increasing stress. Moreover, the intensity of $B_x$ drops monotonically with increasing stress (for this reason we refer to it as $D_x$ for large tensile stress). Data shown in Supplementary Figure 6 and additional collected for different values of the field $F_p$ applied to the actuators were averaged to obtain a single spectrum for each value of $F_p$. To highlight the evolution of the fine structure, such spectra were shifted along the energy axis using the $B_y$ as a reference. The result is shown in Supplementary Figure 7(a), (b). The same procedure was used for another QD (QD2) [Fig. 4(b-d) of the main text], which shows fully consistent behavior. Similar data for still another QD (QD3) are shown in Supplementary Figure 7(c), (d).

Overall, we see very similar behavior for all QDs. In addition, the high energy component $B_z$ is consistently observed only in QDs contained in narrow stripes and not in large membranes. It should be noted that - especially for QD3 - a "full darkening" of the $D_x$ line is not achieved within the applied range of $F_p$ values. We qualitatively attribute this observation to the presence of prestress with main axes away from the [100] and [010] directions, consistent with the large FSS observed at low $F_p$.

Lastly, the polarization orientation of the neutral exciton in QDs embedded in stripe membranes with increasing magnitude of electric field (uniaxial stress) is plotted in Supplementary Figure 7(e). Consistent results are obtained with QDs embedded in macroscopic membranes: the polarization orientations show initially (no intentional uniaxial stress applied) quite random direction, and become perpendicular to the pulling direction ([100] direction) with increasing uniaxial stress. All these results provide compelling evidence that uniaxial stress can be used to deterministically set the orientation of the transition dipoles in a QD, consistent with the theory.

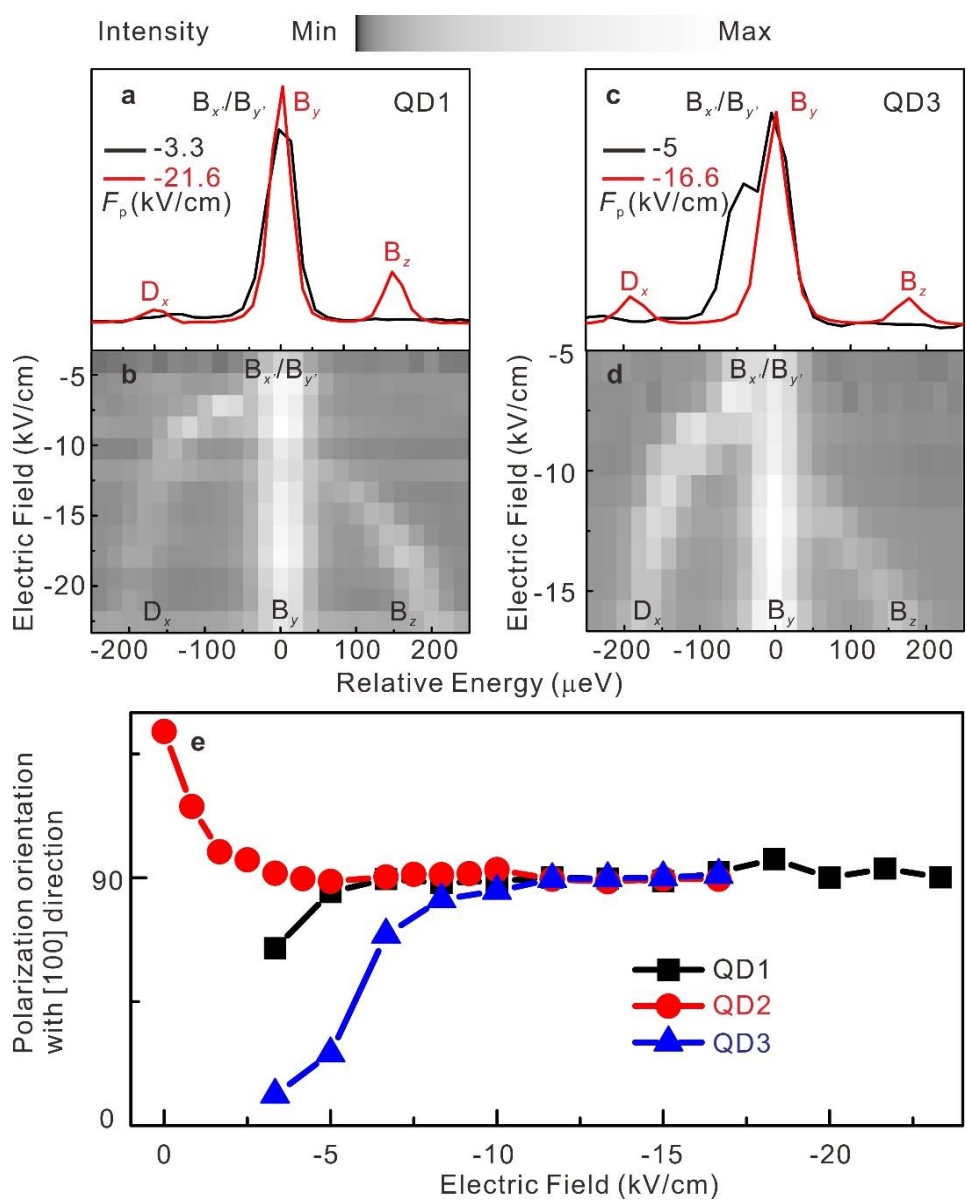

**Supplementary Figure 7**. Evolution of fine structure of the neutral exciton confined in different QDs for increasing magnitude of electric field applied (uniaxial stress). (a), (b) for QD1, (c), (d) for QD3. Similar plots for QD2 are shown in the main text. (e) Evolution of the neutral exciton polarization orientation for different QDs for varying magnitude of electric field applied to the actuator (uniaxial stress).

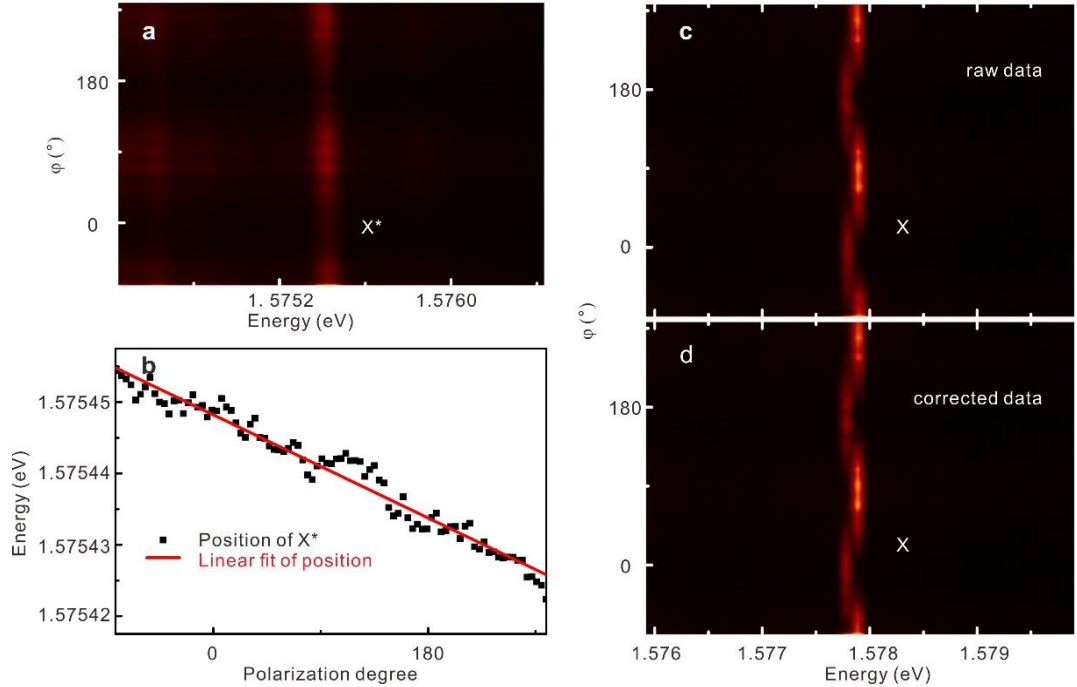

**Supplementary Figure 8**. Illustration of the procedure used to compensate piezo-creep after data acquisition. (a) Typical polarization-resolved PL spectra of a trion (marked as X*) emission. (b) Energy position of the trion emission as a function of polarization angle [obtained from Gaussian fitting of the spectra shown in (a)] and its linear fit. (c), (d) the raw and corrected polarization resolved PL spectra of the neutral exciton (marked as X) emission of the same dot.

**Supplementary Note 7. Data correction for piezo creep-compensation**

Piezo-creep is a quite common phenomenon in ferroelectric materials and typically results in a logarithmically-varying strain at a fixed applied electric field[12]. As mentioned in the main text piezo-creep was observed during the measurements, especially at large $F_p$. This problem, which is not significant for small fields and can be fully compensated by an active feedback,[10,11] was addressed here by combining waiting times of about 10-20 min after setting a new piezo-voltage value for new polarization-resolved measurements and data correction after acquisition. To illustrate the procedure, we take the QD shown in Fig. 3 in the main text with applied electric field of -10 kV/cm as an example. Generally the PL signal of the trion is chosen as a reference due to its rather high intensity, narrow full-width-at-half-maximum (FWHM), and absence of fine-structure, as shown in Supplementary Figure 8(a). Then, through Gaussian curve fit of the trion emission line, the peak position for each polarization angle is plotted, see Supplementary Figure 8(b). For sufficiently long waiting times and relatively short acquisition time for the PL data, the creep-induced shift is approximately linear. A slope of the linear fit of the peak position vs polarization angle is used to slightly shift the polarization-resolved PL spectra. Supplementary Figure 8(c), (d) show the raw and corrected polarization resolved PL spectra of neutral exciton emission, respectively.

**Supplementay Note 8. Definition of quantization axis based on the atomistic empirical-pseudopotential-method (EPM) and the configuration-interaction (CI)**

The HGS confined in the QD was calculated with the EPM+CI methods and projected to the eigenstates of the angular-momentum-projection operator $J_n=\mathbf{J}\cdot\mathbf{n}$, with $\mathbf{n}=(\cos\vartheta, 0, \sin\vartheta)$ in the *x-z* plane for increasing tensile stress. (The eigenstates can be obtained from those of the $J_z$ operator usind the transformations provided in Supplementary Note 9). The result for the projection $|\langle HGS|HH_n\rangle|^2$ is graphically shown in Supplementary Figure 9. Large values of the projection – let's say above 0.9 (a typical value of $|\langle HGS|HH_z\rangle|^2$ for conventional Stranski-Krastanow QDs) can be interpreted as the indication that it is meaningful to define a quantization axis along the direction specified by $\mathbf{n}$. Low values indicate instead mixed HH-LH states, for which the total angular momentum is not well defined. The plot clearly shows that a z quantization axis is appropriate for our as-grown QDs, while the quantization axis is oriented along the x direction for large values of strain). The data shown in red in Fig. 1**b** and 1**c** of the main text correspond to vertical scans for $\vartheta=0$ (*z*-axis) and $\vartheta=\pi/2$ (*x*-axis), respectively.

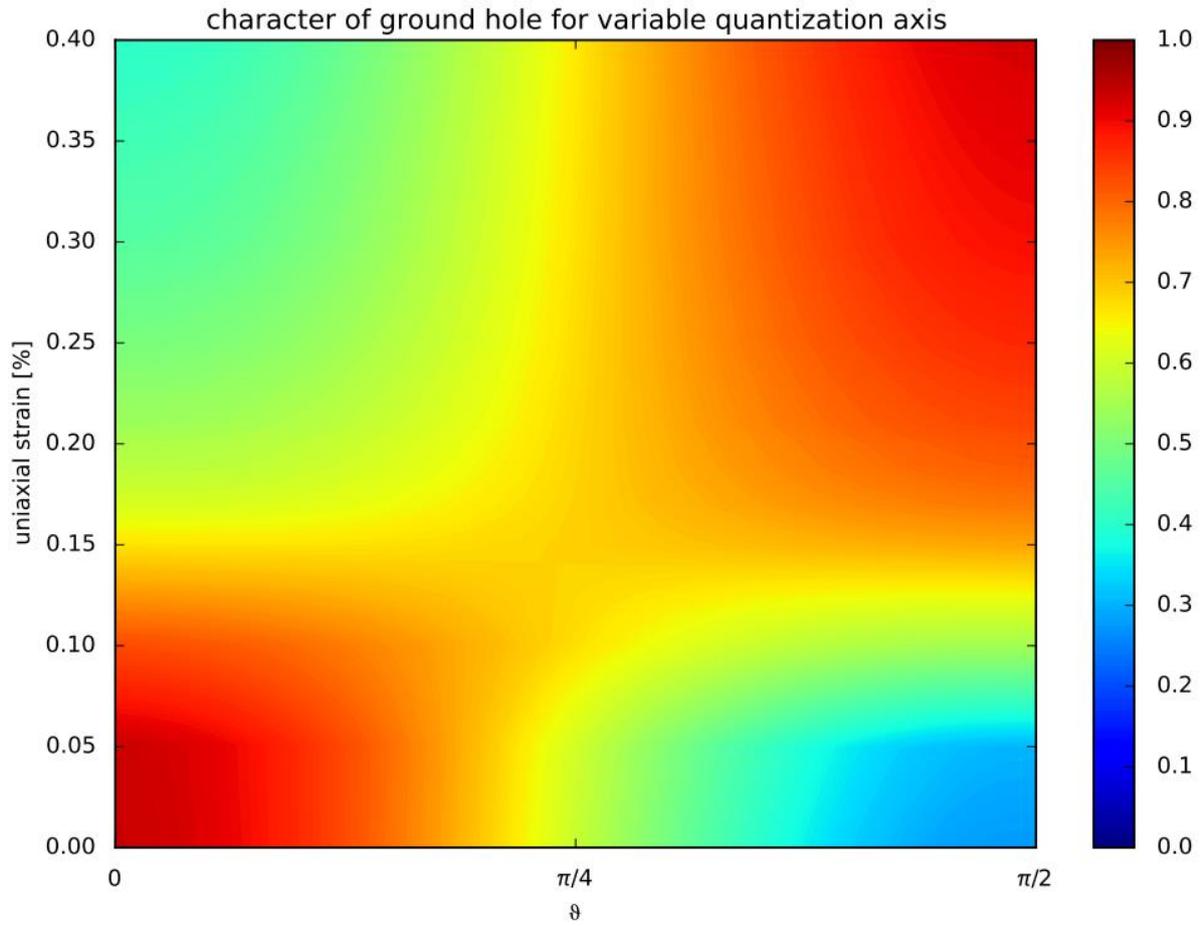

**Supplementary Figure 9.** Value of the projection of the HGS on the HH$_n$ state for different values of strain. Large values of the projection indicate that it is meaningful to define a quantization axis along the direction specified by the unit vector **n**=(cos$\vartheta$, 0, sin$\vartheta$) in the *x-z* plane ($\vartheta$=0 corresponds to the *z*-axis while $\vartheta$=π/2 corresponds to the *x* axis).

**Supplementary Note 9. k.p model: bulk limit**

To illustrate the effect of uniaxial stress on the quantization axis of our QDs we have used **k.p** theory, which is well suited to describe the band structure of semiconductors close to a specific point of the Brillouin zone (here the zone center, i.e. the Γ point).

Since the CB is energetically well separated from the VBs, we have neglected coupling and treated separately the CB and the VBs. The dispersion of the 6 VBs and the effect of strain are described by the Luttinger-Kohn and Pikus-Bir Hamiltonians. For the periodic part of the Bloch wave functions $|u_i\rangle, i = 0 \ldots 8$ for electrons in the CB and holes in the HH, LH, SO bands at the Γ point we use the convention of Ref.[18]:

$$|u_{CB}^1\rangle = iS\uparrow; \qquad |u_{CB}^2\rangle = iS\downarrow$$

$$|u_{HH}^1\rangle = -\frac{1}{\sqrt{2}}(X+iY)\uparrow;$$

$$|u_{LH}^1\rangle = -\frac{1}{\sqrt{6}}[(X+iY)\downarrow -2Z\uparrow]; \quad |u_{LH}^2\rangle = \frac{1}{\sqrt{6}}[(X-iY)\uparrow +2Z\downarrow] \tag{4}$$

$$|u_{HH}^2\rangle = \frac{1}{\sqrt{2}}(X-iY)\downarrow$$

$$|u_{SO}^1\rangle = \frac{1}{\sqrt{3}}[(X+iY)\downarrow +Z\uparrow]; \quad |u_{SO}^2\rangle = \frac{1}{\sqrt{3}}[(X-iY)\uparrow -Z\downarrow]$$

Here $X$, $Y$, $Z$ have the same angular dependence as $(\cos\varphi\sin\theta, \sin\varphi\sin\theta, \cos\theta)$ and can be expressed as proper linear combination of the spherical harmonics with angular momentum quantum number $l=1$ (see Ref. [18]) and S is spherically symmetric ($l=0$). The vertical arrows stand with spin up and spin down states with respect to the z quantization axis. In the main text we have called the $X$, $Y$, $Z$ states as $|X_i\rangle$. By using this basis, the Hamiltonian reads:

$$H = -\begin{pmatrix} -C & 0 & 0 & 0 & 0 & 0 & 0 & 0 \\ 0 & -C & 0 & 0 & 0 & 0 & 0 & 0 \\ 0 & 0 & P+Q & -S & R & 0 & -\frac{S}{\sqrt{2}} & \sqrt{2}R \\ 0 & 0 & -S^+ & P-Q & 0 & R & -\sqrt{2}Q & \sqrt{\frac{3}{2}}S \\ 0 & 0 & R^+ & 0 & P-Q & S & \sqrt{\frac{3}{2}}S^+ & \sqrt{2}Q \\ 0 & 0 & 0 & R^+ & S^+ & P+Q & -\sqrt{2}R^+ & -\frac{S^+}{\sqrt{2}} \\ 0 & 0 & -\frac{S^+}{\sqrt{2}} & -\sqrt{2}Q^+ & \sqrt{\frac{3}{2}}S & -\sqrt{2}R & P+\Delta & 0 \\ 0 & 0 & \sqrt{2}R^+ & \sqrt{\frac{3}{2}}S^+ & \sqrt{2}Q^+ & -\frac{S}{\sqrt{2}} & 0 & P+\Delta \end{pmatrix} \tag{5}$$

with:

$$C = E_v + E_g + C_k + C_\varepsilon, \quad C_k = \frac{\hbar^2 k^2}{2m^*}, \quad C_\varepsilon = a_c(\varepsilon_{xx} + \varepsilon_{yy} + \varepsilon_{zz})$$

$$P = E_v + P_k + P_\varepsilon, \quad P_k = \frac{\hbar^2 k^2}{2m}\gamma_1, \quad P_\varepsilon = -a_v(\varepsilon_{xx} + \varepsilon_{yy} + \varepsilon_{zz})$$

$$Q = Q_k + Q_\varepsilon, \quad Q_k = \frac{\hbar^2}{2m}\gamma_2(k^2 - 3k_z^2), \quad Q_\varepsilon = -\frac{b}{2}(\varepsilon_{xx} + \varepsilon_{yy} - 2\varepsilon_{zz})$$

$$R = R_k + R_\varepsilon, \quad R_k = \frac{\hbar^2}{2m}(\gamma_2(k_x^2 - k_y^2) + 2i\sqrt{3}\gamma_3 k_y k_x), \quad R_\varepsilon = \frac{\sqrt{3}b}{2}(\varepsilon_{xx} - \varepsilon_{yy}) - id\varepsilon_{zy}$$

$$S = S_k + S_\varepsilon, \quad S_k = \frac{\hbar^2}{2m}\gamma_3\sqrt{3}(k_x k_z - ik_y k_z), \quad S_\varepsilon = -d(\varepsilon_{xz} - i\varepsilon_{yz}) \tag{6}$$

$x$, $y$, and $z$ correspond to the [100], [010] and [001] crystal directions, respectively; $E_g$ is the energy bandgap and $E_v$ the energy of the HH and LH valence bands at the $\Gamma$ point in absence of strain; the parameters $\gamma$ are the Luttinger parameters, $m^*$ is the electron effective mass and $a_c$, $a_v$, $b$, $d$ are the deformation potentials. The values of these parameters were taken from Ref. [19].

By diagonalizing the Hamiltonian for a fixed value of **k** we obtain 8 eigenvalues $\lambda_i$ (each at least twice degenerate) and the corresponding eigenvectors $|u_v^i\rangle$ for the valence bands (in general a mixture of HH, SO and LH with different spins) and $|u_c^j\rangle$ for the two conduction bands. In general, the eigenstates corresponding to the VBs can be written as:

$$|u_v^i\rangle = \sum_{m=0}^{8} a_m^i |u_m\rangle \qquad (7)$$

Note that in the model used above the coefficients corresponding to the CB are zero. Since the basis states are linear combinations of the *X, Y, Z* states with spin-up and spin-down, the eigenstates of the Hamiltonian can also be expressed as a linear combination of such states, as stated in the main text.

The model can be used to calculate the band-structure in proximity of **k**=**0** as illustrated in Figs. 1(b-c) of the main text. As an additional remark we would like to add that the nature ("heavy" or "light") of the highest energy VB can be qualitatively understood in a tight-binding picture. A $HH_x$ state has only contributions from atomic orbitals lying in the *y-z* plane and no contribution from $p_x$ orbitals (similar to a $HH_z$ state having contributions from atomic orbitals in the *x-y* plane and no contribution from $p_z$ orbitals, see form of $|u_{HH}^{1,2}\rangle$ above). Upon tension along the *x* direction the overlap between the $p_x$-like atomic orbitals decreases along the "stretched" direction, weakening the bonds involved. In contrast, the overlap between the $p_y$ and $p_z$-like atomic orbitals increases (to a lesser extent than the reduced overlap because the relative displacement along *x* (strain component $\varepsilon_{xx}$) is much larger than $\varepsilon_{yy}$). A decreased/increased overlap leads to a drop/increase of the energy associated with the bond (due to Heisenberg's principle). Following this argument it is clear that the energy of the $HH_x$ state increases under tension, while that of the $LH_x$ state drops with respect to the unstrained situation. This simple picture allows us also to understand the origin of the anisotropic effective mass of the two bands. The $HH_x$ state has no $p_x$ orbitals, making the motion of electron motion along the *x* direction "difficult". The effective mass of such electrons is thus "large". In the perpendicular direction motion is instead "easy" because of the presence of $p_y$ and $p_z$ orbitals with improved overlap. For the $LH_x$ state we have instead "easy" motion along *x* (although with an increased effective mass compared to the unstrained case) and "difficult" in the perpendicular direction (because the LH$_x$ state consists mostly of p$_x$ orbitals). Similar arguments can be applied for the compressive case. It should be noted that the topmost band (LH$_x$) is characterized by an extremely light effective mass in this case, since compression along *x* produces a pronounced improvement of the overlap between p$_x$ atomic orbitals).

We now provide an additional argument to illustrate that the quantization axis can be rotated by strain. For sufficiently large strain, we can neglect the effects produced by confinement. In addition, because of the relatively large value of the spin-orbit coupling-constant Δ we can limit our attention to the topmost 4 bands, which are described by the following 4×4 block of the Hamiltonian in Eq. 1:

$$H_{4\times 4} = -\begin{pmatrix} P+Q & -S & R & 0 \\ -S^+ & P-Q & 0 & R \\ R^+ & 0 & P-Q & S \\ 0 & R^+ & S^+ & P+Q \end{pmatrix} \quad (8)$$

For compressive biaxial stress in the *x-y* plane (which we have mentioned in the main text), the non-zero components of the strain tensor are $\varepsilon_{xx} = \varepsilon_{yy} < 0$ and $\varepsilon_{zz} = -\frac{2C_{12}}{C_{11}}\varepsilon_{xx} \sim -0.9\varepsilon_{xx} > 0$.

With the same basis used for the Hamiltonian, the operator for the *z* component of the total angular moment of an electron in the HH or LH bands is given by:

$$\hat{J}_z = \hbar \begin{pmatrix} \frac{3}{2} & 0 & 0 & 0 \\ 0 & \frac{1}{2} & 0 & 0 \\ 0 & 0 & -\frac{1}{2} & 0 \\ 0 & 0 & 0 & -\frac{3}{2} \end{pmatrix} \quad (9)$$

It is easy to see that this operator commutes with the Hamiltonian at the Γ point, since both *S* and *R* are zero in the case of $\varepsilon_{xx} = \varepsilon_{yy}$. This means that it is possible to choose the eigenstates of $H_{4\times 4}$ and $\hat{J}_z$ simultaneously. In other words, the eigenstates have well defined $J_z$.

In the case of uniaxial stress along *x* we have $\varepsilon_{zz} = \varepsilon_{yy}$ and $\varepsilon_{xx} = -\frac{C_{11}+C_{12}}{C_{12}}\varepsilon_{zz} \sim -3.2\varepsilon_{zz}$. We see from this expression, that uniaxial stress produces a strain configuration, which is much more anisotropic than biaxial stress, as mentioned in the main text. Now *S* is still 0 (because there is no shear strain), but $R$ does not vanish. Since the operator for the component $x$ of the total angular moment is given by:

$$\hat{J}_x = \hbar \begin{pmatrix} 0 & \frac{\sqrt{3}}{2} & 0 & 0 \\ \frac{\sqrt{3}}{2} & 0 & 1 & 0 \\ 0 & 1 & 0 & \frac{\sqrt{3}}{2} \\ 0 & 0 & \frac{\sqrt{3}}{2} & 0 \end{pmatrix} \qquad (10)$$

it is easy to see that now $[\hat{J}_x, H_{4\times 4}]=0$, i.e. that the quantization axis is parallel to the $x$ axis.

## SSupplementary Note 10. Inclusion of vertical confinement and calculation of mixing

Because of the flat morphology of the studied QDs, the main effect of confinement stems from the vertical ($z$) direction. We included the effect of vertical confinement in our **k·p** model following two approaches. The simplest is to emulate the effect of confinement by adding a fixed biaxial compression, as described in the text. After diagonalization of the Hamiltonian, the eigenstates are projected either to the (HH, LH, SO)$_z$ states or to the (HH, LH, SO)$_x$ states to obtain the degree of mixing for the $z$ or the $x$ quantization axis (shown in Fig. 1(e,f)). The (HH, LH, SO)$_x$ states are obtained from the conventional (HH, LH, SO)$_z$ states using the transformations provided in Ref. [18]:

$$\begin{pmatrix} \uparrow' \\ \downarrow' \end{pmatrix} = \begin{pmatrix} e^{-i\frac{\varphi}{2}}\cos\frac{\vartheta}{2} & e^{i\frac{\varphi}{2}}\sin\frac{\vartheta}{2} \\ -e^{-i\frac{\varphi}{2}}\sin\frac{\vartheta}{2} & e^{i\frac{\varphi}{2}}\cos\frac{\vartheta}{2} \end{pmatrix} \begin{pmatrix} \uparrow \\ \downarrow \end{pmatrix} \qquad (11)$$

$$\begin{pmatrix} X' \\ Y' \\ Z' \end{pmatrix} = \begin{pmatrix} \cos\vartheta\cos\varphi & \cos\vartheta\sin\varphi & -\sin\vartheta \\ -\sin\varphi & \cos\varphi & 0 \\ \sin\vartheta\cos\varphi & \sin\vartheta\sin\varphi & \cos\vartheta \end{pmatrix} \begin{pmatrix} X \\ Y \\ Z \end{pmatrix} \qquad (12)$$

with $\vartheta = 90°, \varphi = 0°$ for the $x$-axis.

For Fig. 1(d-f) we have used a biaxial compression of -120 MPa to obtain results close to those obtained by the EPM method. The Bloch wavefunctions shown in Fig. 1 were obtained by replacing the states $X$, $Y$, and $Z$ with their representation in polar coordinates. The evolution of the Bloch wavefunction of the topmost VB states in prestressed bulk GaAs under uniaxial stress is shown in the provided movie. This was obtained by assuming a fixed biaxial stress $\sigma_{xx} = \sigma_{yy} = -120$ MPa superimposed to a variable uniaxial stress $\sigma_{xx}$ from 0 to 320 MPa.

A more correct approach to include the vertical confinement is to consider a GaAs quantum well (QW) with thickness $h_{QW}$ embedded in $Al_{0.4}Ga_{0.6}As$ barriers. The confined states can be calculated by combining **k·p** theory with the effective mass equation. In this approach, $k_z$ in Eq. 1 is replaced by the operator $-i\frac{\partial}{\partial z}$ and the spatial dependence of the VB-edge energy $E_V(z)$ acts as a potential. For the VB profile we have used simple linear interpolations (Vegard's law) of the values of $E_{v,av}$ for GaAs and AlAs provided in Table C.2 of Ref.[18] while all other parameters are obtained using the Vegard's law on parameters provided in Ref.[19]. The uniaxial stress applied to the structure is varied continuously and the corresponding strain values are assumed to be constant throughout the heterostructures (we used the stiffness constants of GaAs also for the barriers). We solved the QW problem by the finite difference method. The $z$ direction is discretized in N steps and each of the elements appearing in Eq. 1 is replaced by a N×N block. To guarantee that the resulting 6N×6N matrix describing the VBs is Hermitian we used the approximations provided by Ref.[20] for the derivatives (Note that in the second term of Eq. 35 one of the $A(z_{i-1})$ should be replaced by $A(z_{i+1})$). Diagonalization is performed with the LA_EIGENVEC routine (from LAPACK) implemented in IDL6.4.

Supplementary Figure 9 shows the result for the mixing evaluated along the $z$ axis and $x$ axis for $h_{QW}$=4 and 12 nm. (Our QDs have a height between 7 and 9 nm). It is evident that the trends closely resemble those obtained with the simplified "bulk" model with a fixed prestress and the more sophisticated 3D EPM model (see Fig. 1(b-c) of main text). We note that for a QW we have a pure $HH_z$ state in absence of strain. In a lens-shaped QD this is not the case because the lateral confinement and other effects induces some LH-HH mixing, as previously calculated for the dots considered here in Refs.[17,21]. In addition we see that the transition from a $z$- to an $x$-oriented quantization axis becomes smoother when the quantum well thickness decreases. In particular we see that for a thin QW substantial mixing persists under large uniaxial tension. This means that tall QDs are required to achieve full rotation of the axis and reduce residual mixing. In addition, compressive stress present in commonly studied SK dots would further hinder the rotation of the quantization axis.

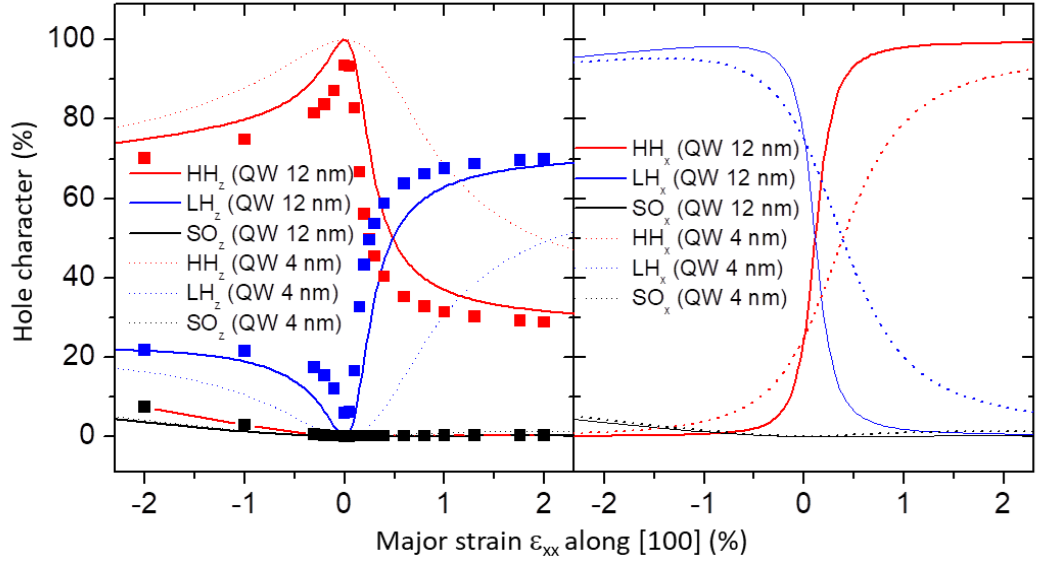

**Supplementary Figure 9.** Calculation of mixing for GaAs/AlGaAs QWs of different thickness. The rotation of the quantization axis is facilitated by a larger QW thickness. This means that tall GaAs QDs are preferable. The results of the EPM calculations shown in Fig. 1 of the main text are reported as symbols.

**Supplementary Note 11. Emission energy shift under uniaxial stress along [100]**

Since we have not determined the strain induced by our actuators experimentally, we can estimate the strain by comparing the observed energy shift to that expected from calculations. Supplementary Figure 10 shows the calculated emission energy of an exciton in a GaAs QD computed by the EPM+CI methods and a simple **k·p** calculation for QWs with thickness ranging from 4 to 12 nm (in this case no excitonic effects were included). While we have linear shifts for sufficiently large tensile strains, the behavior is strongly non-linear at small strains and in the compressive regime. This effect, which is seen both in the EPM and **k·p** calculations, stems from an anticrossing of the two uppermost VBs, as illustrated in Supplementary Fig. 11, where we have used again the bulk Hamiltonian of Eq. 1. (To emulate the behaviour of a QW, we have shifted the CB terms by 52.8 meV (calculated confinement energy for electrons in a 8 nm QW), the diagonal HH terms ($P+Q$) by 9.1 meV and added a fixed energy shift of 10 meV to the diagonal terms corresponding to LH.

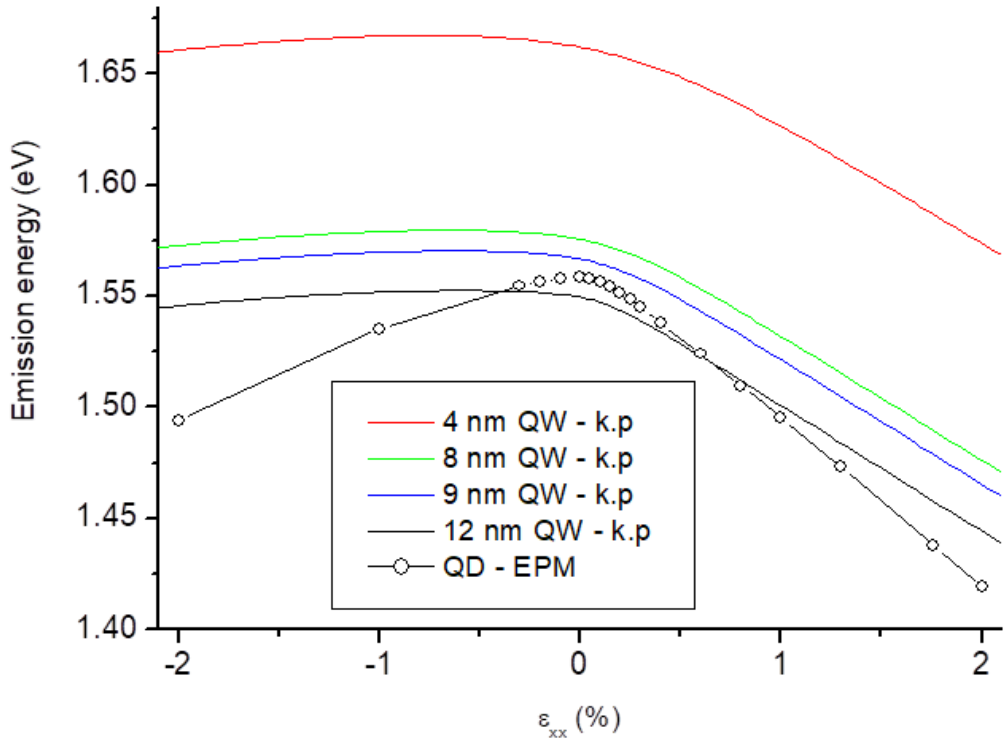

**Supplementary Figure 10.** Calculated emission energy of a GaAs QD via the EPM and CI methods and comparison with the emission energy of QWs with different thickness calculated with the **k·p** method.

The EPM and **k·p** calculations follow qualitatively the same trends. However the EPM results shows enhanced slopes both under tension and compression. Although the reason is not yet clear, we can estimate that an energy shift of 100 meV (see Fig. 2c of the main text) corresponds to a major strain of about 1.5-1.7%.

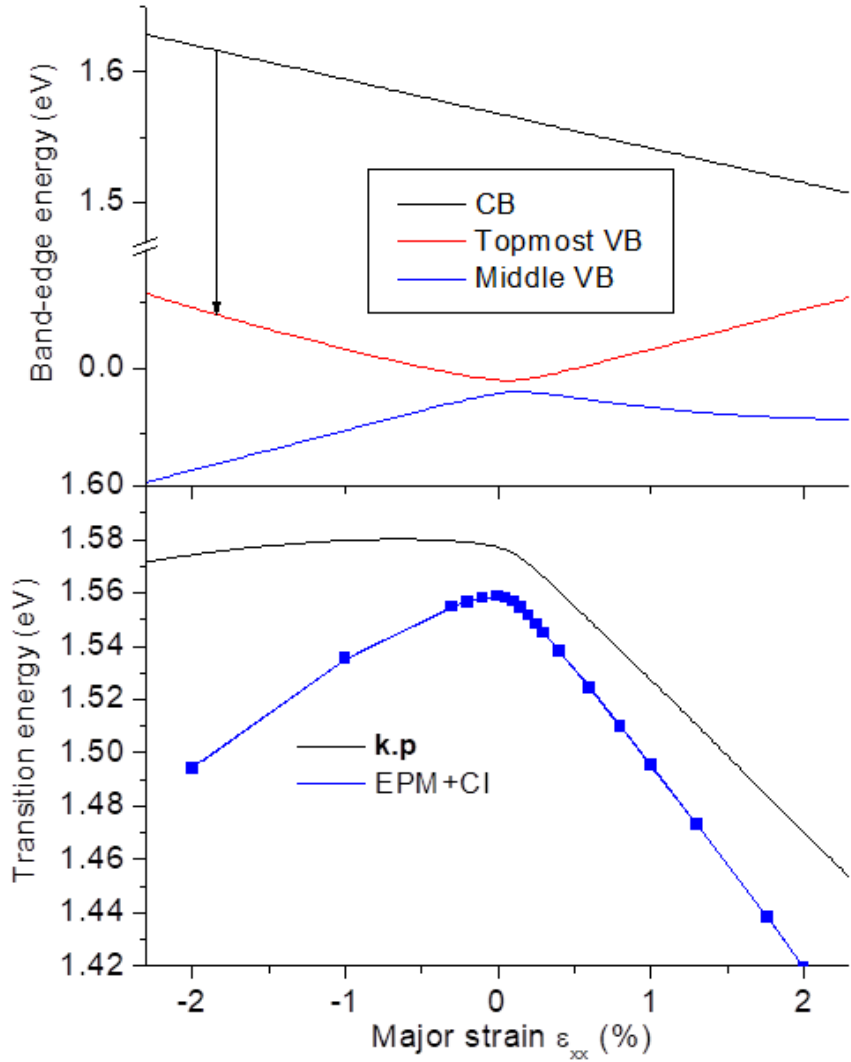

**Supplementary Figure 11.** Illustration of the origin of the non-linear energy shift of the QD emission.

**Supplementary Note 12. Evolution of selection rules under uniaxial stress**

The change of quantization axis upon uniaxial stress changes substantially the selection rules for dipole-allowed transitions. Both theoretically and experimentally we have seen that initially bright transitions turn dark and initially dark transitions turn bright.

By using the experimentally measured radiative lifetime $\tau$ of bright excitons in unstrained QDs (with $HH_z$ ground state)[22] we can estimate the transition rates $\tau^{-1}$ for transitions with polarization along the axes of the cubic cell from both the EPM and $\mathbf{k}\cdot\mathbf{p}$ calculations as shown in Fig. 5(a) of the main text. For the sake of completeness we show in Supplementary Figure 12 the calculated transition strengths for a QW. At 0 strain the HGS has dominant $HH_z$ character according to the 3D EPM calculations (or pure $HH_z$ character according to the 1D calculations for a 12-nm thick QW). Under tensile strain we have the transition to a $HH_x$ HGS, which does not couple to *x*-polarized light. The transition rate for the *x*-polarized emission is slightly higher than for the initially unstrained QD, while the *z*-polarized transition is slightly weaker. In other words the overall oscillator strength, which is initially equally shared between the in-plane transition-dipoles is redistributed in a slightly asymmetric fashion under uniaxial tension. A direct comparison with the experiment would require time-resolved measurements of temporal decay.

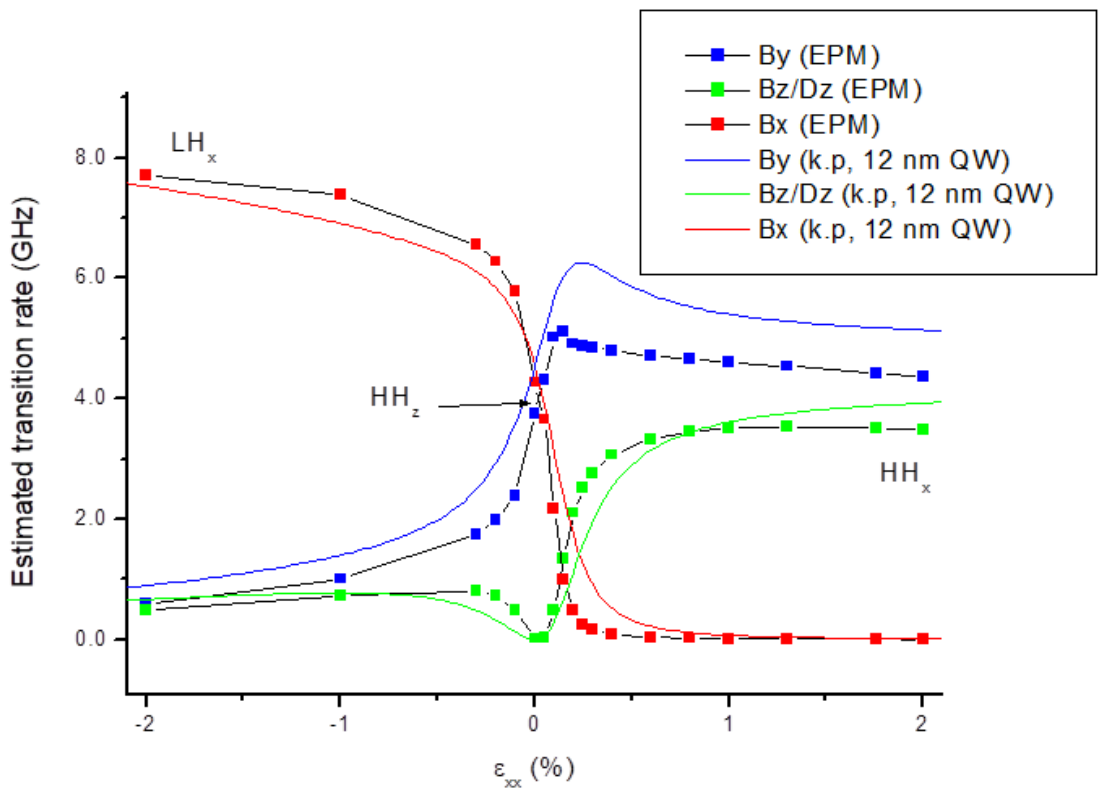

**Supplementary Figure 12.** Estimated transition rates for light polarized parallel to the *x, y,* and *z* direction.

Under compression (not investigated experimentally yet, but very interesting for integrated-photonics applications), almost the whole oscillator strength is transferred to the *x*-polarized emission. The effect is robust, as it is reproduced both by the 3D and 1D simulations and stems from the properties of bulk GaAs (shown in Fig. 5(a) of the main text).

**Supplementary Reference**